\documentclass[12pt]{article}
\usepackage{amsmath,amssymb,amsfonts,color,graphicx,color,soul}
\usepackage{epsfig}
\usepackage{blindtext}
\usepackage[utf8]{inputenc}
\usepackage{caption}
\usepackage{subcaption}
\usepackage{capt-of}
\usepackage{tabu}
\usepackage{hyperref}
\usepackage{csquotes}
\usepackage{amsmath,amsfonts,amssymb}
\usepackage{graphicx}
\usepackage{latexsym,amstext,amscd}
\usepackage{setspace}
\usepackage{epic}
\usepackage{mathrsfs}
\usepackage[T1]{fontenc}
\usepackage[T1]{fontenc}
\usepackage[utf8]{inputenc}

\usepackage[backend=biber,
  sorting=none, 
  url=true,                  
  hyperref=true,
  style=numeric-comp]{biblatex}
\evensidemargin \oddsidemargin
\allowdisplaybreaks

\usepackage[bottom]{footmisc}
\usepackage[affil-it]{authblk}
\usepackage[usenames]{xcolor}
\usepackage[english]{babel}
\usepackage{dcolumn}
\newcommand{\nn}{\nonumber}

\newcommand{\be}{\begin{eqnarray}}
\newcommand{\ee}{\end{eqnarray}}

\newcommand{\Lagr}{\mathcal{L}}

\newcommand{\Zd}{{\mathcal{Z}}_2}

\title{Scalar and gauge sectors in the 3-Higgs Doublet Model under the $S_3$ symmetry }

\author[1]{M. G\'omez-Bock\footnote{melina.gomez@udlap.mx}}

\author[2]{M. Mondrag\'on\footnote{myriam@fisica.unam.mx}}
\author[2]{A. P\'erez-Mart\'inez \footnote{adrianapema7.5@gmail.com}}
\affil[1]{\small Universidad de las Am\'ericas Puebla, UDLAP. Ex-Hacienda Sta. Catarina M\'artir, Cholula, Puebla, M\'exico.}
\affil[2]{Instituto de F\'isica, Universidad Nacional Aut\'onoma de
  M\'exico\\    Apdo. Postal 20-364, M\'exico 01000 D.F., M\'exico.}

\begin{document}

\maketitle
 \begin{center}{\it {Dedicated to the memory of Prof. Alfonso Mondragón,\\ whose love of physics has been an inspiration to so many.
 }}
 \end{center}

\begin{abstract}
We analyse the Higgs sector of an $S_3$ model with three Higgs doublets and no CP violation.  After electroweak breaking there are nine physical Higgs bosons, one of which corresponds to the Standard Model one. We study the scalar and gauge sectors of this model, taking into account the conditions set by the minimisation and stability of the potential. We calculate the masses, trilinear and quartic  Higgs-Higgs, and Higgs-gauge couplings.  We consider two possible alignment scenarios, where only one of the three neutral scalars has couplings to the gauge bosons and corresponds to the SM Higgs, and whose trilinear and quartic couplings  reduce exactly to the SM ones. We also obtain numerically the allowed parameter space for the scalar masses in each of the alignment scenarios.   We use the calculated trilinear and quartic couplings to find the analytical structure of the one-loop neutral scalar mass matrix, without fermionic contributions.   We show that it is possible to have a compact mass spectrum where the contribution to the oblique parameters might be small. We explore some scenarios for the loop contributions to the neutral scalar masses.

\end{abstract}

\section{Introduction}
 
 The discovery the Higgs boson with a mass of $125$ GeV \cite{Aad:2012tfa,Chatrchyan:2012ufa}, and the
experimental study of its properties,  will be relevant to gain a deeper understanding of the flavour problem and of ways to address it.

The organization of the fermions into generations or families may signal a possible underlying structure in elementary particles, although its origin or nature is not yet understood. On
the other hand,  the Standard Model (SM) 
Higgs mechanism \cite{Higgs:1966ev,Gunion:1989we}, which is indispensable to understand the origin of the 
masses of gauge bosons and fermions, 
sheds no light on the flavor structure or the difference in the masses
of the fundamental fermions.

The flavour structure of fermions has been the subject of a great
amount of research throughout the years. In view of the fact that the only difference between
generations in the fermionic sector are  the masses of the
particles, the most direct or even natural way to propose a
flavour structure is through the mass generation mechanism, the Higgs sector.

One possibility to understand the flavour nature behind the SM is to construct an extended scalar sector with a flavour symmetry, where the SM is embedded. Multi-Higgs extensions of the SM, with and without extra symmetries, have been extensively studied, some diverse examples are given in  \cite{Flores:1982pr,Ginzburg:2004vp,Branco:2005em, Nishi:2007nh,Osland:2008aw,Ferreira:2008zy,Olaussen:2010aq,Kubo:2012ty, Yagyu:2016whx,Bento:2018fmy,deMedeirosVarzielas:2019rrp} (for reviews on two Higgs Doublet Models (2HDM) and multi-Higgs models see \cite{Branco:2011iw,Ivanov:2017dad}). 
Discrete symmetries have been extensively studied in this context, both at low and at high energies (for  reviews of models with discrete symmetries see  \cite{Ishimori:2010au,Ishimori:2012zz,Altarelli:2010gt,King:2013eh}). 
Since  these models in general require 
 the addition of more Higgs fields, 
the phenomenological consequences in all sectors, like allowed extra processes and couplings, have to be analysed, some examples can be found in  \cite{Moretti:2015tva,Camargo-Molina:2017klw,Akeroyd:2019mvt}. 
Restrictions are placed on the models by confronting their phenomenology with the experimental results, in this case the ones of ATLAS \cite{Aad:2019mbh} and CMS \cite{Sirunyan:2018koj}. The Higgs sector is thus crucial to determine the viability and prospects of each model. Prime examples of this procedure are the Minimal Supersymmetric Standard Model (MSSM) and 2HDM (see for instance, \cite{Gunion:1984yn,Carena:2002es,Heinemeyer:2014uoa} and \cite{Gunion:2002zf,Branco:2011iw, Bento:2017eti}, respectively).

The permutation group of three
objects $S_3$, with three Higgs doublets, has been proposed already a
long time ago \cite{Pakvasa:1977in,Pakvasa:1978tx,Harari:1978yi,Derman:1979nf,Barnhill:1984gf} as a
natural extension of the SM, even before all the quarks were
discovered or the mass of the neutrinos established.  Since then, the $S_3$ symmetry has been extensively studied
in different contexts, both in the quark \cite{Mondragon:1998yw, Mondragon:1998gy, Lavoura:1999dn, Mondragon:1999jt, Mondragon:2007nk, Canales:2013cga, Hernandez:2013hea, Das:2015sca, Ge:2018ofp, Gomez-Izquierdo:2018jrx} and lepton sectors \cite{Ma:1991eg, Kubo:2003iw, Chen:2004rr, Jora:2006dh, Felix:2006pn, Mondragon:2007af, Mondragon:2007jx, Jora:2009gz, Dicus:2010iq, Meloni:2010aw, Canales:2012dr,    Dias:2012bh,Hernandez:2014lpa, Felix-Beltran:2014xpa, Ma:2014qra, Cruz:2017add, Xing:2019edp, Garcia-Aguilar:2020vsy}, due to its simplicity and predictivity, as well as  in the scalar sector \cite{Kubo:2004ps, Das:2014fea, Barradas-Guevara:2014yoa, Hernandez:2015dga, Barradas-Guevara:2015rea, Emmanuel-Costa:2016vej, Haber:2018iwr, Kuncinas:2020wrn} where more predictions arise. More recently, there have been also studies of dark matter candidates in  models with  $S_3$ symmetry  \cite{Chakrabarty:2015kmt,Machado:2016hca,Espinoza:2018itz,Mishra:2019keq,Espinoza:2020qyf,Khater:2021wcx}. 

In particular, the 3 Higgs doublet model with $S_3$ symmetry (which we will refer here as S3-3H) \cite{Kubo:2003iw}, has led to very interesting results in the fermionic sector.  
 In the quark sector, it was shown that it is possible to obtain the Fritzsch  and the Nearest Neighbour Interaction (NNI) textures \cite{Canales:2013cga}, thus fitting the CKM matrix.
In the leptonic sector it was found that the S3-3H model can also reproduce the $V_{PMNS}$ matrix and predicts a non-vanishing $\theta_{13}$ reactor mixing angle, and some flavour changing neutral
currents and contributions to $g-2$ were calculated \cite{Mondragon:2007jx,Mondragon:2007af,Mondragon:2007nk,Canales:2012dr}.
 In \cite{Espinoza:2018itz}, a version 
of the S3-3H with an extra inert Higgs doublet was analysed (S3-4H), with the interesting result that it is possible to have a good dark matter (DM) candidate,
coming from the inert sector and satisfying also the Higgs bounds. The indirect prospects of detection of this DM candidate have been studied in \cite{Espinoza:2020qyf}.  Although there has been extensive work in models with $S_3$ symmetry and three Higgs doublets in different contexts, the phenomenological implications in the Higgs sector have not been fully explored. Our motivation to analyse more closely the scalar sector of the S3-3H model concerns the fact that it is in this sector where novel experimental signatures can be found; also the results in the scalar sector will have an impact and allow for a deeper analysis of the fermionic sector.

The conditions for stability and symmetry breaking in the general three Higgs doublet model (3HDM) have been studied in \cite{Maniatis:2014oza}. Assuming an extra discrete symmetry reduces greatly the number of free parameters, in particular the case of the S3-3H potential  was already analysed in \cite{Kubo:2004ps}, although requiring a soft breaking of the discrete symmetry. The vacuum stability of the S3-3H scalar potential, without soft breaking of $S_3$,  was studied in \cite{EmmanuelCosta:2007zz,Beltran:2009zz}, and the mass structure of the scalar bosons was analysed in  \cite{Das:2014fea,Barradas-Guevara:2014yoa}. In \cite{Das:2014fea} it was found that there is a residual $\mathcal{Z}_2$ symmetry after the electroweak symmetry breaking (EWSB) in the Higgs potential, and the  corresponding charges for the scalars under this symmetry  were given. The conditions for having spontaneous CP violation in this potential were  presented in \cite{Emmanuel-Costa:2016vej}.

In here, we keep the model
as simple as possible, by not assuming an explicit breaking of the
flavour symmetry or adding extra flavons.  We calculate the scalar masses, and  the trilinear and quartic Higgs self-couplings and Higgs-gauge boson couplings. We consider two possible alignment scenarios for the SM-like Higgs boson, where only one of the three neutral scalars has couplings to the vector bosons (one is always decoupled due to the $\mathcal{Z}_2$ symmetry). We use a geometrical parameterization in spherical coordinates, which allows us to express the mixing of the  vacuum expectation values ({\it vevs}) of the Higgs fields in the $S_3$ singlet and doublet irreducible representations, in terms of one angle ($\theta$) in our expressions. We scan the parameter space, taking into account the unitarity and stability conditions, and the SM Higgs boson mass constraints,  in each of the two alignment scenarios.  

Some of the trilinear scalar couplings have been obtained in \cite{Barradas-Guevara:2014yoa}, nevertheless we found differences with their results.  Mainly in \cite{Barradas-Guevara:2014yoa}, the $\mathcal{Z}_2$ symmetry is not exhibited, whereas we find it explicitly in our calculations, consistent with the $\mathcal{Z}_2$ residual symmetry reported in \cite{Das:2014fea}. As an additional result, we find  that in each of the alignment limits, where only the SM-like Higgs couples to the vector bosons,  the trilinear and quartic  couplings reduce exactly to the SM ones. 

We find the  expression for the  neutral scalar mass matrix at one-loop, where  all the scalar and gauge contributions derived from the calculated trilinear and quartic couplings are taken into account. Although it reduces to a structure similar to the 2HDM mass matrix, due to the residual ${\mathcal Z}_2$ symmetry, the presence of an extra neutral scalar in our case, $h_0$,  allows to distinguish between the models. It is possible to find values for the parameter $\theta$ and the scalar masses where the off-diagonal term of the one-loop  neutral scalar mass matrix vanishes, thus minimising the radiative corrections.  We give two examples of such spectra, one with light and another one with heavier masses.  The latter one fulfils the conditions  that can make  
the contributions to the oblique parameters to  be small or even vanish \cite{Grimus:2008nb,Hernandez:2015rfa}.

The paper is organized as follows: in the next section we describe the model, and how the $S_3$ symmetry  acts on the Higgs electroweak doublets,  giving the structure and characteristics of the Higgs potential in the S3-3H model. In Section 3, we parameterize the vacua and rotate to the Higgs basis, to express our results in terms of physical parameters. We then calculate the tree level masses and explore numerically the two different alignment scenarios. Then, in Section 4, we calculate the Higgs-Higgs couplings and  the Higgs-gauge bosons couplings; we present the ones involving neutral scalars in this section, and we complete  with the pseudoscalars and charged scalar couplings in the Appendix. We also analyse the structure of  the one-loop neutral scalar  mass matrix.
Finally, we present a summary and the conclusions of our work.

\section{The S3-3H model scalar sector}
\label{S3Lag-sec}

We will discuss briefly here how the $S_3$ symmetry is implemented in the scalar sector of the model. 
The $S_3$ group is the smallest non-Abelian discrete group, it corresponds
to the rotations and reflections that leave invariant an equilateral
triangle, or equivalently, to the permutations of three objects.  It
has 
three irreducible representations (irreps):  a symmetric singlet
${\bf 1}_S$, an anti-symmetric singlet ${\bf 1}_A$,  and  a doublet $\bf
2$ \cite{Ishimori:2010au}. 

The multiplication rules among the irreducible
representations are as follows
\begin{align}\label{rules}
&{\bf 1}_{S}\otimes {\bf 1}_{S}={\bf 1}_{S},\quad {\bf 1}_{S}\otimes {\bf 1}_{A}={\bf 1}_{A},\quad {\bf 1}_{A}\otimes {\bf 1}_{S}={\bf 1}_{A},\quad {\bf 1}_{A}\otimes {\bf 1}_{A}={\bf 1}_{S},\nn\\&
{\bf 1}_{S}\otimes {\bf 2}={\bf 2},\quad {\bf 1}_{A}\otimes {\bf 2}={\bf 2},\quad {\bf 2}\otimes {\bf 1}_{S}={\bf 2},\quad {\bf 2}\otimes {\bf 1}_{A}={\bf 2};\nn\\
&\begin{pmatrix}
a_{1} \\ 
a_{2}
\end{pmatrix}_{{\bf 2}}
\otimes
\begin{pmatrix}
b_{1} \\ 
b_{2}
\end{pmatrix}_{{\bf 2}} = 
\left(a_{1}b_{1}+a_{2}b_{2}\right)_{{\bf 1}_{S}} \oplus  \left(a_{1}b_{2}-a_{2}b_{1}\right)_{{\bf 1}_{A}} \oplus	
\begin{pmatrix}
a_{1}b_{2}+a_{2}b_{1} \\ 
a_{1}b_{1}-a_{2}b_{2}
\end{pmatrix}_{{\bf 2}}. 
\end{align}
$S_3$ has 6 subgroups: the trivial group, the whole group, three ${\mathcal Z}_2$ subgroups (which correspond to the reflections over the axes of symmetry of the triangle), and a ${\mathcal Z}_3$ subgroup. 

We will consider here three $SU(2)$ electroweak (EW) Higgs  doublets, i.e. two more
than in the Standard Model. We   will  assign here two of the Higgs
EW doublets to the $\bf 2$ irrep of $S_3$ and the third one 
to the symmetric singlet ${\bf 1}_S$, but in this work we will
concentrate only on the scalar sector, so our results are general
for any scalar potential of this type, irrespective of the assignment
for the fermionic sector. In ref.~\cite{Kubo:2003iw} an extension
of the SM was considered, with three $SU(2)$ Higgs doublets plus  three
right-handed neutrinos, we refer to this model as S3-3H.
In the fermionic 
sector of the S3-3H model, prior to EWSB, the first two generations of quarks and leptons,
as well as two of the Higgs doublets, 
were assigned to the $S_3$ doublet irrep, and the third generation of
fermions and one Higgs electroweak doublet, to the
symmetric singlet irrep. After EWSB all the fields are mixed, giving rise to a specific texture for the mass matrices of quarks and leptons. 

 \subsection{The S3-3H model scalar potential}
\label{S3VHiggs-sec}
The terms in the potential
are the ones that preserve the discrete $S_3$ permutational symmetry,
as reported in \cite{Derman:1979nf, Kubo:2004ps}.  The most general
Higgs potential invariant under the $SU(3)_c \times SU(2)_L \times U(1)_Y
\times S_3$ in the symmetry adapted basis, according to the multiplication rules (\ref{rules}), is given as, 

{\normalsize\begin{eqnarray}\label{Hpotential}
V&=&\mu^2_1 \left( H^\dagger_1 H_1 + H^\dagger_2 H_2\right) + \mu^2_0 \left(H^\dagger_s H_s \right) + \frac{a}{2}\left(H^\dagger_s H_s \right)^ 2 + b\left(H^\dagger_s H_s \right)\left( H^\dagger_1 H_1 + H^\dagger_2 H_2\right) \nonumber \\
&&+\frac{c}{2}\left( H^\dagger_1 H_1 + H^\dagger_2 H_2\right)^ 2 + \frac{d}{2} \left( H^\dagger_1 H_2 - H^\dagger_2 H_1\right)^ 2 + ef_{ijk}\left(\left(H^\dagger_s H_i\right)\left(H^\dagger_j H_k\right) + h.c.\right)\nonumber\\
&& +f\left\lbrace \left( H^\dagger_s H_1\right)\left( H^\dagger_1 H_s\right) + \left( H^\dagger_s H_2\right)\left( H^\dagger_2 H_s\right) \right\rbrace  + \frac{g}{2} \left\lbrace \left( H^\dagger_1 H_1 - H^\dagger_2 H_2\right)^2 + \left( H^\dagger_1 H_2 + H^\dagger_2 H_1\right)^2 \right\rbrace \nonumber\\ 
&& + \frac{h}{2}\left\lbrace \left( H^\dagger_s H_1\right)\left( H^\dagger_s H_1\right)+ \left( H^\dagger_s H_2\right)\left( H^\dagger_s H_2\right) + \left( H^\dagger_1 H_s\right)\left( H^\dagger_1 H_s\right) + \left( H^\dagger_2 H_s\right)\left( H^\dagger_2 H_s\right) \right\rbrace; \label{c3e5}
\end{eqnarray}}

\noindent where $f_{112} = f_{121} = f_{211} = -f_{222} = 1$. This same
potential has also been analysed in
Refs.~\cite{EmmanuelCosta:2007zz,Beltran:2009zz,Barradas-Guevara:2015rea,Das:2014fea}
without CP violation, and in Ref.~\cite{Emmanuel-Costa:2016vej} 
with spontaneous CP violation.  We will only consider here the case without CP violation, i.e. solutions with real {\em vevs.}

As already mentioned, we will assign two of the Higgs doublets, $H_1$ and $H_2$ to the doublet  irrep of $S_3$ ${\mathbf 2}$, and the third one, $H_S$, to the symmetric singlet irrep ${\bf 1}_S$.

In terms of complex fields we express them as
\begin{eqnarray}
\label{SU2doublets}
H_1 =  \frac{1}{\sqrt{2}}\begin{pmatrix}   \phi_1 + i\phi_4 \\
 					\phi_7 + i\phi_{10}  \end{pmatrix},\ \   H_2 = \frac{1}{\sqrt{2}}\begin{pmatrix}   \phi_2 + i\phi_5 \\
 					\phi_8 + i\phi_{11}  \end{pmatrix},\ \ 
 					H_s = \frac{1}{\sqrt{2}} \begin{pmatrix}   \phi_3 + i\phi_6 \\
 					\phi_9 + i\phi_{12}  \end{pmatrix}. 
\end{eqnarray}

In order to simplify  the calculations, we introduce the following variables as  was done in \cite{EmmanuelCosta:2007zz,Beltran:2009zz}
\begin{eqnarray}
\begin{matrix}
x_1 = H^\dagger_1 H_1, & x_4 = Re(H^\dagger_1 H_2), & x_7 = Im(H^\dagger_1 H_2), \\
  x_2= H^\dagger_2 H_2, & x_5 = Re (H^\dagger_1 H_s), & x_8 = Im(H^\dagger_1 H_s),\\
  x_3 = H^\dagger_s H_s, & x_6 = Re (H^\dagger_2 H_s), & x_9= Im(H^\dagger_2 H_s).
 \end{matrix} \label{xnewvar}
\end{eqnarray}
As an example, we show here explicitly some  of the real terms of the scalar fields in the potential, with the appropriate normalization factors 
\begin{eqnarray}
x_1& =& H^\dagger_1 H_1
=\frac{1}{2}(\phi_1^2+\phi_4^2+\phi_7^2+\phi_{10}^2),
\nonumber\\
x_4 &=&Re(H_1^\dagger H_2)
=\frac{1}{2}(\phi_1\phi_2+\phi_4\phi_5+\phi_7\phi_8+\phi_{10}\phi_{11}),
\nonumber\\
x_7 &= &Im(H_1^\dagger H_2)
=\frac{1}{2}(\phi_1\phi_5-\phi_4\phi_2+\phi_7\phi_{11}-\phi_{10}\phi_{8}).
\end{eqnarray}
\noindent
Hence, using (\ref{xnewvar}) into (\ref{Hpotential}), the Higgs potential is expressed as:
\begin{eqnarray}
V &=&\mu^2_1 (x_1 +x_2) + \mu^2_0 x_3 + \frac{a}{2} x_3^2 + b(x_1 + x_2 )x_3 + \frac{c}{2} (x_1 +x_2)^2  \nonumber \\
&&- 2dx_7^2 + 2e\left[ (x_1 -x_2)x_6 + 2x_4 x_5 \right] + f(x_5^2 + x_6^2 + x_8^2 +x_9^2) \nonumber \\
&&+ \frac{g}{2} \left[(x_1 - x_2 )^ 2 + 4x_4^2 \right]  + h(x_5^2 + x_6^2 -x_8^2 - x_9^2).  \label{Hpotx}
\end{eqnarray}

\noindent From this general potential we have ten free parameters, before EWSB.

\subsection{The normal minimum}
\label{NVmin}

In order to have a consistent Higgs potential, it is necessary to check
that it is stable, i.e. bounded from below, and that it respects
perturbative unitarity.  These requirements impose constraints on the
potential's parameters.  This analysis has already been done in
\cite{Das:2014fea}, and we use their expressions for the unitarity and
stability bounds in here.

A study of the stability of the different minima for a general Higgs potential of this
kind can be found in \cite{EmmanuelCosta:2007zz}, they point out
the existence of three types of minima or stationary points.
In here, we will consider the EWSB using
the natural choices of conservation of electric and CP charges,
implying that only the real part of the neutral fields will acquire
{\it vevs}, we will refer to this as the {\it normal minimum}. 
Thus, only the real parts of each one of the doublets will acquire
non-zero vacuum expectation values.  Expressed in terms of the field
components of $H_1, H_2, H_s$, Eq.(\ref{SU2doublets}) we have
\begin{eqnarray}
\langle\phi_7\rangle = v_1, \langle\phi_8\rangle = v_2, \langle\phi_9\rangle = v_3, \langle\phi_i\rangle = 0, \ \ i\neq 7, 8, 9,\label{phivevs}
\end{eqnarray}
\noindent this adds two more free parameters to the model, as they should satisfy the condition
 \begin{equation}
 \sqrt{v_1^2 + v_2^2 + v_3^2} = v = 246 \ \text{GeV}~.
 \end{equation}

The extreme point conditions for the potential are given by
\begin{eqnarray}
\frac{\partial V}{\partial v_i} = 0 \longleftrightarrow \frac{\partial V}{\partial x_j}\frac{\partial x_j}{\partial v_i} = 0, \label{minimal}
\end{eqnarray}
with $i = 1, 2, 3;$ $j=1,2,...,9$. These conditions express the tree level tadpole equations as
\begin{eqnarray}
0&=&[2\mu^2_1 +(b+f+h)v^2_3 + (c+g)(v^2_1+v^2_2)]v_1 + 6ev_{1}v_{2}v_{3},\label{c3e11}\\
0&=&[2\mu^2_1 +(b+f+h)v^2_3 + (c+g)(v^2_1+v^2_2)]v_2 +3e(v^2_1 - v^2_2)v_3, \label{c3e12}\\ 
0&=&[2\mu^2_0 +(b+f+h)(v^2_1+v^2_2)+ av^2_3]v_3 +e(3v^2_1 - v^2_2)v_2. \label{c3e13}
\end{eqnarray}
\noindent
These equations reduce further the original twelve free parameters relating two of them as
\begin{eqnarray}
v_1^2 = 3 v^2_2. \label{v1v2}
\end{eqnarray}
Another possible solution that satisfies these equations would be
$e=0$ \cite{Kubo:2003iw,Beltran:2009zz}, which implies the presence of a Goldstone
boson due to a residual $SO(2)$ symmetry, but this scenario will not
be considered for the present work. The general minima of this potential, both real and complex,  have been studied in \cite{Emmanuel-Costa:2016vej}, with emphasis on the complex vacua.  In here, we will consider only in detail the case (\ref{v1v2}), with $v_1=+\sqrt{3}v_2$, where after EWSB there is a residual ${\mathcal Z}_2$ symmetry. This residual symmetry  corresponds to  one of the ${\mathcal Z}_2$ subgroups of $S_3$, namely, a reflection over one of the symmetry axes of the triangle.  In a similar fashion, the solution $v_1=-\sqrt{3}v_2$  has also a residual ${\mathcal Z}_2$ symmetry, which is another one of the subgroups of $S_3$. In this latter case the invariance is under the reflection over the opposite axis of symmetry as  the positive solution.  This negative solution leads exactly to the same results for the masses and couplings as the positive solution.

\section{Tree level Higgs masses and physical basis}
\label{Hmasses}

In order to get the tree level  masses of the Higgs bosons, it is
necessary to 
diagonalize the $12\times 12$ matrix resulting from taking the second
derivatives of the potential 
 \begin{eqnarray}
 (\mathcal{M}^2_H)_{ij} = \left.\frac{\partial^2 V}{\partial\phi_i \partial\phi_j}\right|_{\langle \phi_i\rangle}, \label{c3e22}
 \end{eqnarray}
 with $i, j = 1, ..., 12$. Due to the symmetry of the model, the mass matrix consists of four diagonal blocks,
each one a $3\times 3$ Hermitian and symmetric matrix. The Higgs mass matrices of the S3-3H model have been reported previously in \cite{Barradas-Guevara:2014yoa,Das:2014fea},  nevertheless our results
differ from \cite{Barradas-Guevara:2014yoa}  
by a factor of two in the Higgs couplings, because
we have included the normalization factors $1/\sqrt{2}$ in the Higgs doublets. In this work, we will study the  general case with $e\neq 0$, with a new parameterization which allows us to compare directly with the SM when we include the complete scalar couplings and scalar-gauge couplings. 

Since we assume no CP violation, we obtain three $3\times 3$ Hermitian matrices, one  for the charged scalars $\mathbf{M}^2_C$, one for the neutral scalars $\mathbf{M}^2_S$, and one for the pseudoscalar bosons masses $\mathbf{M}^2_A$.

 The matrix elements of the charged Higgs masses in terms of the potential parameters  are given as
\begin{eqnarray}
 \mathbf{M}^2_C =\begin{pmatrix}
 c_{11} & c_{12}& c_{13} \\ c_{21} & c_{22} & c_{23} \\ c_{31} & c_{32 }& c_{33}  \end{pmatrix}, \label{c3e24}
 \end{eqnarray}
\noindent with  the elements of the symmetric mass matrix given as 
\begin{eqnarray}
&c_{11} = -v_3 [2 ev_2+\frac{v_3}{2} (f+ h)]- g v_2^2, &
 c_{12} =  \sqrt{3} v_2 (e v_3+ g v_2),\notag\\
& c_{13} = \sqrt{3} v_2[ e v_2+\frac{v_3}{2} (f+ h)], &
 c_{22} = -v_3 [4 e v_2+\frac{v_3}{2} (f+ h)]-3 g v_2^2, \notag \\
& c_{23} = v_2 [e v_2 + \frac{v_3}{2}(f + h)], &
 c_{33} = -\frac{2 v_2^2 [2 e v_2+v_3 (f+ h)]}{v_3}.
 \label{eq:elementsymm}
\end{eqnarray}

  The mass  matrix for neutral scalars is given by
  \begin{eqnarray}
 \mathbf{M}^2_S =\begin{pmatrix}
 s_{11} & s_{12}& s_{13} \\ s_{21} & s_{22} & s_{23} \\ s_{31} & s_{32} & s_{33}  \end{pmatrix}, \label{c3e25}
 \end{eqnarray}
 \noindent where, the elements of the scalar symmetric mass matrix are
\begin{eqnarray}
 &s_{11}= 3 v_2^2 (c+g), 
 & s_{12}=  \sqrt{3} v_2[ v_2(c+g)+3 e v_3], \notag \\
 &s_{13} =  \sqrt{3} v_2 [v_3 (b+f+ h)+3 e v_2],
 &s_{22}= v_2 [v_2 (c+g)-6 e v_3], \notag \\
& s_{23} =  v_2 [3 e v_2 + (b + f +  h) v_3], 
 &s_{33}= \frac{ \left(a v_3^3- 4e v_2^3\right)}{v_3}.
 \label{eq:elementneutral}
\end{eqnarray}

 For the pseudoscalar mass matrix we find
 \begin{eqnarray}
 \mathbf{M}^2_A =\begin{pmatrix}
 a_{11} & a_{12}& a_{13} \\ a_{21} & a_{22} & a_{23} \\ a_{31} & a_{32} & a_{33}  \end{pmatrix}, \label{c3e26}
 \end{eqnarray}
where each of the elements of the symmetric matrix are given as
\begin{eqnarray}
 & a_{11} = - \left(v_2^2 (d+g)+2e v_2 v_3+h v_3^2\right), &
 a_{12} = \sqrt{3} v_2 (v_2 (d+g) + e v_3), \notag \\
& a_{13} = \sqrt{3} v_2 (e v_2+ hv_3), &
 a_{22} = -3 v_2^2 (d+g)-4 e v_2 v_3- h v_3^2, \notag \\
&a_{23} = v_2 (e v_2 +  h v_3), &
 a_{33} = -\frac{4 v_2^2 (e v_2+ h v_3)}{v_3}.
\end{eqnarray}
 
 \subsection{Geometrical parameterization of the vacua and the Higgs masses}
 We will rewrite the {\it vevs} in spherical coordinates, as it was done in \cite{Espinoza:2017ryu,
 Espinoza:2018itz}
\begin{eqnarray}
v_1=v\cos\varphi\sin\theta, & v_2= v\sin\varphi\sin\theta, & v_3= v\cos\theta.
\end{eqnarray}

The use of this spherical parameterization is helpful to visualize the relation among the {\it vevs}. The angle $\theta$ gives the amount of mixing between the {\it vev} of the singlet ($v_3$) and the {\it vevs} of the doublet ($v_1, v_2$).  
We express the relations between $v_1$ , $v_2$, and $v_3$ in terms of two angles as: 
\begin{equation}
\tan\varphi = \frac{v_2}{v_1},\quad\quad \tan\theta  = \frac{v_2}{v_3\sin\varphi}.
\label{mixanglesvev}
\end{equation}

Moreover, the minimization condition of the potential (\ref{v1v2}), provides an extra constraint for the relation between $v_1$ and $v_2$ {\it i.e.} it fixes also the value of $\varphi$. We assume all the {\it vevs} to be real and positive  (otherwise we should consider a phase between two {\it   vevs}) implying $\varphi=\pi/6$, then $\tan\varphi=\frac{1}{\sqrt{3}}$, thus we get 
\begin{eqnarray}
\tan\varphi=1/\sqrt{3}&\Rightarrow& \sin\varphi=\frac{1}{2},\ \ \  \  \ \ \ \cos\varphi= \frac{\sqrt{3}}{2}, \label{phi}\\
\tan\theta=\frac{2v_2}{v_3}&\Rightarrow& \sin\theta=\frac{2v_2}{v}, \ \ \  \,  \ \ \ \cos\theta= \frac{v_3}{v}.\label{angtheta}
\end{eqnarray}
\noindent 

The usual form for the rotation matrix $R_i$, to obtain the mass matrix and physical states is given as:
 \begin{eqnarray}
 [\mathcal{M}^2_{diag}]_{I} = R^{T}_I {\bf M}^2_I R_I, \ \ \ \  I=S,A,C.\label{DiagHiggsMass}
 \end{eqnarray}
 For the Higgs bosons we take the sub-indices $I=S, A, C$ to refer to the neutral, pseudoscalar and charged Higgs bosons respectively. The rotation matrix is the product of two rotations, i.e., $R_I= A\, B_I$, where 

\begin{eqnarray}
 A= \begin{pmatrix}
\cos\delta & -\sin\delta & 0 \\ \sin\delta & \cos\delta & 0 \\ 0 & 0 & 1 
\end{pmatrix}, \ \ \ \ \ B_I =  \begin{pmatrix}
\cos\gamma_I & 0 & \sin\gamma_I \\ 0 & 1 & 0 \\ -\sin\gamma_I & 0 & \cos\gamma_I 
\end{pmatrix},
\end{eqnarray}
then
\begin{eqnarray}
R_I=
\begin{pmatrix}
\cos\gamma_I\cos\delta & -\sin\delta & \sin\gamma_I\cos\delta \\ \cos\gamma_I\sin\delta & \cos\delta & \sin\gamma_I\sin\delta \\ -\sin\gamma_I & 0 & \cos\gamma_I  
\end{pmatrix}.
\end{eqnarray}

The rotation matrix $R_{A, C}$, which diagonalizes  $\mathcal{M}^2_A$ and $\mathcal{M}^2_C$, will transform the fields leading to the Goldstone states. They are given as follows:
 
\begin{eqnarray}
R_{A, C} =  
\begin{pmatrix}
\cos\gamma_{A, C}\cos\delta & -\sin\delta & \sin\gamma_{A, C}\cos\delta \\ \cos\gamma_{A, C}\sin\delta & \cos\delta & \sin\gamma_{A, C}\sin\delta \\ -\sin\gamma_{A, C} & 0 & \cos\gamma_{A, C}  
\end{pmatrix}  =\begin{pmatrix}
\frac{\sqrt{3} v_2}{v} & -\frac{1}{2} & -\frac{\sqrt{3}v_3}{2 v}  \\ 
\frac{v_2}{v} &  \frac{\sqrt{3}}{2} & -\frac{v_3}{2 v} \\
\frac{v_3}{v} & 0 & \frac{2v_2}{v} 
\end{pmatrix}. \label{c3e28}
\end{eqnarray} 
Therefore, we can see that $\cos\gamma_{A, C}=  \frac{2v_2}{v} $, $\sin\gamma_{A, C} = -\frac{v_3}{v}$, $\sin\delta = \frac{1}{2}$, and $\cos\delta = \frac{\sqrt{3}}{2}$. If we compare with (\ref{phi}) and (\ref{angtheta}) we can see that $\delta =  \varphi $ and $\gamma_{A, C}= \frac{3\pi}{2} +  \theta $. We will reparameterize the matrices in terms of the angles $\theta$  and $\varphi$, so that the matrices take the following form
 \begin{eqnarray}
R_{A, C} =  
\begin{pmatrix}
\sin\theta\cos\varphi & -\sin\varphi & -\cos\theta\cos\varphi \\ \sin\theta\sin\varphi & \cos\varphi & -\cos\theta\sin\varphi \\ \cos\theta & 0 & \sin\theta
\end{pmatrix}  =\begin{pmatrix}
\frac{\sqrt{3} v_2}{v} & -\frac{1}{2} & -\frac{\sqrt{3}v_3}{2 v}  \\ 
\frac{v_2}{v} &  \frac{\sqrt{3}}{2} & -\frac{v_3}{2 v} \\
\frac{v_3}{v} & 0 & \frac{2v_2}{v} 
\end{pmatrix}. \label{c3e28-2}
\end{eqnarray}

Thus, we obtain the respective masses at tree level for the pseudoscalar and charged Higgs bosons as 
 \begin{eqnarray}
m^2_{A_1} &=&-v^2\left[ (d+g)\sin^2\theta +\frac{5}{4}e\sin2\theta +h\cos^2\theta\right], \label{c3e44}\\
m^2_{A_2} &=& - v^2(\frac{e}{2}\tan\theta + h)~, \label{c3e45}
\end{eqnarray} 
 \begin{eqnarray}
 m^2_{H_1^\pm} &=& -\frac{v^2}{4}\left[5e\sin2\theta + 2(f+h)\cos^2\theta  + 4g\sin^2\theta\right], \label{c3e46}\\
 m^2_{H_2^\pm} &=&- \frac{v^2}{2} \left[e\tan\theta + (f+h)\right]. \label{c3e47}
 \end{eqnarray}
From these expressions it can be seen that all masses are proportional to $v$, with their values  determined by an interplay  of the self-couplings and $\theta$.

For the diagonalization of the mass matrix of the neutral scalar bosons $\mathcal{M}^2_S$, we will have the following rotation matrix:
\begin{eqnarray}
R_{S} =  
\begin{pmatrix}
\cos\gamma_{S}\cos\delta & -\sin\delta & \sin\gamma_{S}\cos\delta \\ \cos\gamma_{S}\sin\delta & \cos\delta & \sin\gamma_S\sin\delta \\ -\sin\gamma_{S} & 0 & \cos\gamma_{S}  
\end{pmatrix} . 
\end{eqnarray}
In terms of the parameters of the potential, considering also the spherical parameterization, we have
\begin{eqnarray}\label{MelemSPh}
 M_a^2&= &\left[(c+g) v^2\sin^2\theta +\frac{3}{4} ev^2\sin2\theta\right],\notag\\
 M_b^2 & =&  \left[3ev^2\sin^2\theta + (b + f + h)v^2\sin2\theta\right],\notag\\
 M_c^2&=& av^2\cos^2\theta - \frac{1}{2} e v^2\tan\theta\sin^2\theta ,
 \end{eqnarray} 
where the mixing angle $\alpha$ is
\begin{equation} \label{scalargamma}
  \tan 2\alpha=-\frac{M^2_b}{M_a^2 - M_c^2} ~. 
\end{equation}
The rotation to obtain the mass matrix  directly from the interaction basis is given as
\begin{eqnarray} 
R_S = \begin{pmatrix}
\frac{\sqrt{3}
\left( M_a^2 - M_c^2 + Z_M\right)}{ 2 \sqrt{(M_b^2)^2 + \left(  M_a^2 - M_c^2 + Z_M\right) ^2}}  & -\frac{1}{2} & -\frac{\sqrt{3} M_b^2}{ 2\sqrt{(M_b^2)^2 + \left(  M_a^2 - M_c^2 + Z_M\right) ^2}} \\ 
\frac{
M_a^2 - M_c^2 + Z_M}{ 2\sqrt{(M_b^2)^2 + \left(  M_a^2 - M_c^2 + Z_M\right) ^2}} & \frac{\sqrt{3}}{2}  &  -\frac{M_b^2}{ 2\sqrt{(M_b^2)^2 + \left(  M_a^2 - M_c^2 + Z_M\right) ^2}} \\
\frac{M_b^2}{ \sqrt{(M_b^2)^2 + \left(  M_a^2 - M_c^2 + Z_M \right) ^2}} & 0 & \frac{
M_a^2 - M_c^2 + Z_M}{ \sqrt{(M_b^2)^2 + \left(  M_a^2 - M_c^2 + Z_M\right) ^2}}
\end{pmatrix}~, \label{c3e29}
\end{eqnarray}
where 
$$Z_M=\sqrt{(M^2_b)^2+(M^2_a-M^2_c)^2}~.$$ 
Using again  Eqs. (\ref{scalargamma}) and.~(\ref{phi})  we get  
\begin{eqnarray}
\tan\gamma_S = -\frac{M_b^2}
{\left( M_a^2 - M_c^2 + Z_M\right)},&
\gamma_{S}= \frac{3\pi}{2} +  \alpha, & \sin\delta= \frac{1}{2} \;\;\text{and}\;\; \cos\delta =  \frac{\sqrt{3}}{2}.
\end{eqnarray}

We will work with the angles $\alpha$  and $\varphi$, for that reason we express the rotation matrix in the following form
\begin{eqnarray}
R_{S}& =&  
\begin{pmatrix}
\sin\alpha\cos\varphi & -\sin\varphi & -\cos\alpha\cos\varphi \\ \sin\alpha\sin\varphi & \cos\varphi & -\cos\alpha\sin\varphi \\ \cos\alpha & 0 & \sin\alpha
\end{pmatrix}.
\end{eqnarray}
\noindent
Then, we may write the scalar Higgs bosons masses as follows:

\begin{eqnarray}
	m_{h_0}^2 &=&-\frac{9}{4} ev^2\sin2\theta, 
	\label{h0masstree}\\
	m_{H_1, H_2}^2 &=&\frac{1}{2}\left[(M_a^2 + M_c^2) \pm \sqrt{(M_a^2 - M_c^2)^2 + (M_b^2)^2}\right],
\label{H12masses}
\end{eqnarray}	
we notice that $e<0$. We see here that the structure of the masses is consistent with the one found in Refs.~\cite{Beltran:2009zz,Das:2014fea}.
The expressions for $m_{H_{1,2}}$ can be written in terms of the parameters of the model as
\begin{eqnarray}
m^2_{H_1}&=&\frac{v^2}{2}\big[ 2ac^2_\theta c^2_\alpha + 2 (c+g)s^2_\alpha s^2_\theta + 4 (b+f+h) s_\alpha s_\theta c_\alpha c_\theta  \nonumber \\
&& + e t_\theta (6s_\alpha c_\alpha s_\theta c_\theta + 3 c^2_\theta s^2_\alpha -s^2_\theta c^2_\alpha )\big],\nonumber \\
&=&\frac{v^2}{4}\left\lbrace a(c_{\alpha-\theta}+c_{\alpha+\theta})^2 + (c+g)(c_{\alpha-\theta}-c_{\alpha+\theta})^2 +  (b+f+h) (c_{2(\alpha-\theta)}-c_{2(\alpha+\theta)}) \right. \nonumber\\
&&\left. + e\, t_\theta \left[c_{2(\alpha-\theta)}-c_{2(\alpha+\theta)} +4s_{\alpha+\theta}s_{\alpha-\theta}+2s^2_{\alpha+\theta})\right] \right\rbrace,
   \label{H1masstree}
\end{eqnarray}
\begin{eqnarray}
m^2_{H_2}&=&\frac{v^2}{2}\big( 2ac^2_\theta s^2_\alpha + 2 (c+g)c^2_\alpha s^2_\theta - 4 (b+f+h) s_\alpha s_\theta c_\alpha c_\theta  \nonumber \\
&&+ e t_\theta (-6s_\alpha c_\alpha s_\theta c_\theta + 3 c^2_\theta c^2_\alpha -s^2_\theta s^2_\alpha) \big) ,\nonumber \\
&=&\frac{v^2}{4}\left\lbrace a(s_{\alpha-\theta}+s_{\alpha+\theta})^2 + (c+g)(s_{\alpha+\theta} -s_{\alpha-\theta})^2 - (b+f+h)( c_{2(\alpha-\theta)}-c_{2(\alpha+\theta)}) \right. \nonumber\\
&&\left. + e\, t_\theta \left[ c_{2(\alpha+\theta)}-c_{2(\alpha-\theta)}+4c_{\alpha+\theta}c_{\alpha-\theta}+2c^2_{\alpha+\theta}\right] \right\rbrace,
	\label{H2masstree}
\end{eqnarray}
here we use the reduced notation for the trigonometric functions:  $s_{x}\equiv \sin x$, $c_{x}\equiv \cos x$ and $t_{x}\equiv \tan x$.

The SM Higgs boson has already been measured \cite{Sirunyan:2018koj,Aad:2019mbh}, and one of the  neutral CP-even Higgs of the model should correspond to it. Thus, it is important to explore the structure of these tree level masses in terms of the self-couplings of the Higgs potential in the interaction basis, Eq.~(\ref{Hpotx}), and also in terms of the mixing angles relating the {\it vevs}, Eqs.~(\ref{mixanglesvev}). The possibility of a neutral Goldstone, a  massless degree of freedom has been reported previously in \cite{Kubo:2003iw,Beltran:2009zz} when $e=0$ is considered, and leads to $v_1 = v_2$.

From Eqs.~(\ref{MelemSPh}), (\ref{h0masstree}) and (\ref{H12masses}), we can reduce the expressions of the masses for specific cases of the parameter $\theta$ which, as we said before, gives the amount of mixing between the {\it vev} of the singlet and the {\it vevs} of the doublets. 
We explore a particular case, for instance $\theta=\pi/4$, where we obtain the neutral CP-even Higgs masses as
\begin{eqnarray}
m^2_{h_0}&=&\frac{-9}{4}e v^2,\label{mh0pi4}\\
    m_{H_{1,2}}^2&=&\frac{v^2}{4}\Bigg[a+c+e+g\pm \sqrt{(-a+c+2e+g)^2+(3e+2(b+f+h))^2}\Bigg] ~\label{mh12pi4},
\end{eqnarray}
where the three masses are proportional to the {\it vev}, $v=246$ GeV, and combinations of the self-couplings.

Setting the {\it vevs} of the doublet or singlet to zero has to be considered from the beginning, to arrive at the appropriate tadpole equations. The case where $v_1=v_2=0$ corresponds to one of the minima found in \cite{Emmanuel-Costa:2016vej,Khater:2021wcx}, which leaves  $\mu_1$  undetermined. In their solution, the three neutral scalar masses  are in principle different from zero, with two of them  degenerate and depending on $\mu_1$.  In our case, when we take the limit $\sin\theta \to 0$, which leads to $v_1,v_2 \to 0$, we get two almost massless scalars. This can be seen from Figure~\ref{figureAB-SM}, as $\tan\theta \to 0$, also $m_{h_0}, m_{H_2} \to 0$. It can also be verified from the structure of the  mass matrices Eqs.~(\ref{H1masstree},\ref{H2masstree}), or by noticing that in this limit also $\alpha \to 0$ (Eq.~(\ref{scalargamma})) and substituting in Eqs.~(\ref{H1masstree},\ref{H2masstree}). The particular choice of $2\mu_1 = -(h +b+f)v_3^2$ in the solution of refs. \cite{Emmanuel-Costa:2016vej,Khater:2021wcx} implies that the two degenerate masses become zero, and the third one coincides with our $m_{H_1} = av^2$. But it is not possible to arrive to the latter condition for $\mu_1$  from our tadpole equations, since we initially have considered  $v_1, v_2 \neq 0$.
Similarly, it is not possible to have exactly $\cos\theta = 0$, or equivalently $v_3=0$ in our case, since Eqs.~(\ref{c3e11}-\ref{c3e13}) are arrived at dividing by $v_3$. If one assumes $v_3=0$ from the beginning, the tadpole equations are different, and they lead to the relationship $v_2^2 = 3 v_1^2$ \cite{Emmanuel-Costa:2016vej}, which is the inverse of the ratio we find between $v_1$ and $v_2$ .

For our particular solution, we will assume none of the {\em vevs} are  zero, and thus there must be an admixture of the doublet and the singlet Higgs fields, which might be relevant when considering the fermionic sector.

In the following section, we will analyse the masses for the general cases of non-zero parameter values. Specifically, for the numerical analysis, we explore the tree level masses for  different  values of the $\theta$ parameter, in the range $0 < \theta < \pi /2$, to keep the {\em vevs} positive.

\subsection{The Higgs basis} 
\label{Higgsbasis}
From the perspective of both EWSB and flavor physics, there is a basis that is particularly useful to compare with the SM or with other of its extensions, the so-called  Higgs basis. It is defined as  the basis in which one Higgs field carries the full {\it vev}, $\phi_{vev} $, and the other Higgs fields $\psi_1, \psi_2 $ are perpendicular to it  \cite{Donoghue:1978cj, Georgi:1979dq, Wells:2009kq, Akeroyd:2016ssd, Yagyu:2016whx, Ivanov:2017dad}.
In order to get the Goldstone bosons, which are needed for the generation of masses of the gauge bosons, we do the usual rotation. 

For multi-Higgs models, the Goldstone bosons are obtained with the same rotation angle for both the pseudoscalars and charged Higgs bosons, and as we found in the previous section $\gamma_A=\gamma_C=\frac{3\pi}{2} + \theta$.  The fields in the Higgs basis are then given by the transformation: 
\begin{eqnarray}
\begin{pmatrix}
\phi_{vev}\\
\psi_1\\
\psi_2
\end{pmatrix}
=R_A^T
\begin{pmatrix}
H_1\\
H_2\\
H_s\\
\end{pmatrix}
=
\begin{pmatrix}
\sin\theta\cos\varphi & \sin\theta\sin\varphi &\cos\theta \\
-\sin\varphi & \cos\varphi & 0\\
-\cos\theta\cos\varphi& -\cos\theta\sin\varphi & \sin\theta
\end{pmatrix}
\begin{pmatrix}
H_1\\
H_2\\
H_s\\
\end{pmatrix}.
\end{eqnarray}
In our case, the rotation matrix takes the following form
\begin{eqnarray}
	\begin{pmatrix}
	\phi_{vev}\\  	\psi_1\\  	\psi_2
	\end{pmatrix}  =  
	\begin{pmatrix}
	\frac{\sqrt{3}v_2}{v} & \frac{v_2}{v} & \frac{v_3}{v} \\    -\frac{1}{2} & \frac{\sqrt{3}}{2} & 0 \\    -\frac{\sqrt{3}v_3}{2v}  & -\frac{v_3}{2v}  & \frac{2v_2}{v}
	\end{pmatrix} 
	\begin{pmatrix}
	H_1 \\  H_2  \\  H_s
	\end{pmatrix} .
	\label{HiggsBasis}
\end{eqnarray}

Then, the electroweak (EW) Higgs doublets in this basis are explicitly given as %
\begin{eqnarray}\label{Hbasis}
 \phi_{vev} =  
	\begin{pmatrix}
	G^{\pm} \\  \frac{1}{\sqrt{2}}(v + \widetilde{H} + iG_0)  
	\end{pmatrix} , \ \ \ \  
	\psi_1 = \begin{pmatrix}
	H_1^{\pm} \\  \frac{1}{\sqrt{2}}( \widetilde{H}_a + iA_1)  
	\end{pmatrix}, \ \ \ \  
	 \psi_2 = \begin{pmatrix}
	H_2^{\pm} \\  \frac{1}{\sqrt{2}}( \widetilde{H}_b + iA_2)  
	\end{pmatrix}.
\end{eqnarray}

The rotation matrix above, (\ref{HiggsBasis}), corresponds to the  matrix built
for the charged  and pseudoscalars Higgs bosons mass eigenstates, denoted by  $H_1^{\pm},~H_2^{\pm}, G^{\pm}, G_0, A_1$ and  $A_2$.
Whereas the neutral part of the Higgs doublets, denoted by $\widetilde{H}, \widetilde{H}_a$ and $\widetilde{H}_b$  do not correspond to their  mass eigenstates, but they are in the Higgs basis. Now, in order to diagonalize the neutral sector we rotate through the angle $\alpha$, obtaining the relationship between the intermediate-basis states in the Higgs basis and the physical states (mass eigenstates) for the neutral  scalars as

\begin{eqnarray}
	\begin{pmatrix}
	\widetilde{H}\\  	\widetilde{H}_a\\  	\widetilde{H}_b
	\end{pmatrix}  =  
	\begin{pmatrix}
	 \cos( \alpha-\theta  )& 0 &\sin(\alpha - \theta  )\\    0 & 1 & 0 \\  -\sin(\alpha - \theta   ) & 0 & \cos(\alpha - \theta )
	\end{pmatrix} 
	\begin{pmatrix}
	H_1 \\  h_0 \\   H_2
	\end{pmatrix}. 
	\label{Hmasseig}
\end{eqnarray}

We can get the neutral  physical states from either  the direct rotation Eq.~(\ref{c3e29}), which transforms the interaction basis to the physical basis (mass basis),  or through a two step rotation, from the interaction basis to the Higgs basis Eq.~(\ref{HiggsBasis}), and then to the physical basis Eq.~(\ref{Hmasseig}). Either way we obtain the scalar Higgs masses given in Eqs.~(\ref{h0masstree}) and (\ref{H12masses}). We can see from expression (\ref{Hmasseig}), that there will be two alignment scenarios: A) When  $\tilde{H}= H_2$  corresponds to the SM-like Higgs boson; B) when we set  $\tilde{H}= H_1$ corresponding to the SM-like Higgs boson. $\tilde{H}_a$ already corresponds to the physical state $h_0$.

The $\mathcal{Z}_2$ parity assignments for the physical and intermediate-basis states are given in Table \ref{tab:my_label}. Notice that  in the alignment limits the intermediate-basis states become the physical states. 

\begin{table}[ht]
    \centering
    \begin{tabular}{|c|c|c|}
\hline
 Neutral scalars   &  Pseudoscalars & Charged scalars\\
\hline
\begin{tabular}{c|c}
       $h_0$  & odd \\
       $\tilde{H}$ &  even \\
        $\tilde{H}_b$ &  even  \\
\hline
\end{tabular}&
\begin{tabular}{c|c}
        $A_1$ & odd    \\
        $A_2$ &  even \\
\end{tabular}&
\begin{tabular}{c|c}
       $H_1^{\pm}$ & odd \\
        $H_2^{\pm}$& even \\
    \end{tabular} \\ 
\hline
\end{tabular}    
\caption{$\mathcal{Z}_2$ parity assignment for the physical states $h_0, A_{1,2}$ and $H^{\pm}_{1,2}$, and the intermediate-basis states $\tilde{H}$, and  $\tilde{H}_b$.  In the alignment limit the last two will  correspond also to the physical states.}
   \label{tab:my_label}
\end{table}

\subsubsection{Gauge-Higgs sector}

In here we examine the scalar kinetic structure of the Lagrangian
through the covariant derivative of the scalar fields.  It is
important to analyse the covariant derivative for scalar doublets in
order to verify not only the electroweak symmetry breaking mechanism (EWSB), {\it i.e.}
the contributions of the {\it vevs} to the gauge boson masses, but also to find the possible
couplings among the Higgs and gauge bosons.
The kinetic terms are taken as usual
\begin{eqnarray}
\mathcal{L}_{kin} = (\mathcal{D}_\mu H_1)^\dagger(\mathcal{D}_\mu H_1)+ 
(\mathcal{D}_\mu H_2)^\dagger (\mathcal{D}_\mu H_2) + (\mathcal{D}_\mu H_s)^\dagger(\mathcal{D}_\mu H_s)\label{kin}~.
\end{eqnarray}

Using the Higgs basis and Eq.~(\ref{Hmasseig}) in order to get the physical states, we  obtain the electroweak gauge bosons masses, $W^{\pm}$ and  $Z^0$, as well as their couplings with the Higgs bosons, including the ones with $A_\mu$, after performing the canonical rotation with the weak angle. 
Using the covariant derivative for the kinetic term, Eq.~(\ref{kin}), in the Higgs basis and  expanding the Higgs field about the vacuum, we get the Lagrangian in the physical basis. We show here some of the terms for illustration, the complete set of explicit couplings are given in next section and in  Appendix \ref{appendixA},
\begin{eqnarray}
\Lagr_{kin} \approx  &&\frac{g^2v^2}{4}W^+_\mu W^{-\mu} + \frac{g^2v\sin(\alpha -\theta)}{2}H_2W^+_\mu W^{-\mu} + \frac{g^2}{4}H_2H_2W^+_\mu W^{-\mu} \nonumber \\ && + \frac{g^2v\cos(\alpha -\theta)}{2}H_1W^+_\mu W^{-\mu} + \frac{g^2}{4}H_1H_1W^+_\mu W^{-\mu} + \frac{g^2}{4}h_0h_0W^+_\mu W^{-\mu}
 + ... + \nonumber \\
&&+ \frac{(g^2 + g^{'2})v^2}{8}Z_\mu Z^\mu + \frac{(g^2 + g^{'2})v\sin(\alpha - \theta )}{4}H_2Z_\mu Z^\mu + \frac{(g^2 + g^{'2})}{8} H_2H_2Z_\mu Z^\mu \nonumber\\ &&+  \frac{(g^2 + g^{'2})v\cos(\alpha - \theta )}{4}H_1Z_\mu Z^\mu + \frac{(g^2 + g^{'2})}{8} H_1H_1Z_\mu Z^\mu  + \frac{(g^2 + g^{'2})}{8}h_0h_0Z_\mu Z^\mu + ... \notag\\
\label{kinHiggs}
\end{eqnarray}
where $H_1,~H_2$ and $h_0$ are the physical states. 
Then, the masses for the EW gauge bosons $W^\pm$ and $Z^0$ are obtained in the usual form
\begin{eqnarray}
m^2_{W^\pm} =  \frac{g^2 v^2}{4}, & \,\,m^2_{Z} = \frac{v^2 }{4}(g^2 + g'^2)~.
\end{eqnarray}

Furthermore, as the model has two different charged Higgs bosons $H_1^{\pm},H_2^{\pm}$, we explicitly
verified that  mixed charged Higgs and gauge bosons couplings do not appear (e.g. $H_1^{+} H_2^-\gamma$) as it should be in order to preserve the ${\cal Z}_2$ symmetry. We show it by calculating explicitly the part of the Lagrangian for the photon (this is exhibited implicitly in \cite{Das:2014fea} as they calculate $H_{SM} \to \gamma \gamma$ through a loop of charged Higgs bosons)
\begin{eqnarray}
\mathcal{L}_{H^+H^-\gamma}=\frac{igg'}{\sqrt{g^2 + g'^2}}\Big( H_1^+\partial_\mu H_1^-- H_1^-\partial_\mu H_1^+ +H_2^+\partial_\mu H_2^--H_2^-\partial_\mu H_2^+\Big)A^\mu~.
\end{eqnarray}
This is the usual expression that appears in other multi-Higgs models with a specific  symmetry \cite{Keller:1986nq}; it coincides exactly with the one for some two Higgs doublet models \cite{Gunion:1984yn}, where a ${\cal Z}_2$ was assumed.  In our case, the ${\cal Z}_2$ is  a subgroup of the original $S_3$, a residual symmetry left over after EWSB, and not imposed separately.

In section \ref{G-Hcoupl} we give the explicit form of the gauge-scalar couplings of the $H_1$ and $H_2$ neutral Higgs bosons of the model, in order to compare with SM couplings scenarios. The rest of the couplings, for the extended scalar sector, are given in Appendix \ref{AppA3}, where we can see the manifestation of the ${\cal Z}_2$ symmetry as it allows only certain couplings. 

\subsection{Higgs masses and scenarios}\label{Hscenarios}

The neutral Higgs boson $h_0$ is decoupled from the other two, due to the residual symmetry ${\cal Z}_2$, as was reported already in \cite{Das:2014fea} (see Table \ref{tab:my_label}).  In the next section, we explicitly calculate all possible tree level couplings among the scalars and also, between the scalars and the gauge bosons. We show that, as expected, due to the ${\cal Z}_2$ symmetry, $h_0$ couples only in even numbers  to the  gauge bosons Eq.(\ref{kinHiggs}).
Also, its trilinear scalar coupling is absent, as we will see in section \ref{scalars}, Eqs.(\ref{h0VV}) and (\ref{3h0}), so it is immediately excluded as the SM-like Higgs boson. Nevertheless, this neutral Higgs boson could be interesting as a possible dark matter candidate, provided it is the lightest particle in the ${\cal Z}_2$ odd sector, and has no couplings to the SM fermions.

The above discussion leaves us with two possible scenarios for which,  either $H_1$ or $H_2$ is aligned to have the mass and couplings of the SM Higgs boson, and the other one would be practically decoupled from the gauge bosons.  We consider both scenarios for the numerical analysis.

Scenario A is defined by setting $H_2$, which has the lower mass among $H_1$ and $H_2$, as the SM-like Higgs boson. We further restrict the  tree level  value of its mass to be in the range 120-130 GeV, taking into account that it will receive radiative corrections. On the other hand, in scenario B  the heavier Higgs boson $H_1$ is taken as  the SM-like one, with its mass restricted to the interval 120-130 GeV, as in the previous scenario. 
For these two scenarios, the alignment means that the SM-like Higgs boson is maximally coupled to the gauge bosons, while the other one is practically decoupled.
 The alignment of the neutral scalar bosons in models with extended scalar sectors, in what would be the equivalent to our scenario A, is discussed in  \cite{Das:2019yad}, as can be derived from Eq.(\ref{Hmasseig}).

 A third, less natural case, would be a non-alignment scenario, where both Higgs bosons would couple equally or similarly to the gauge bosons. This analysis would be more complex, and a way to establish the non-observation of the second neutral Higgs would be needed, we will not consider that possibility here.
 
\subsubsection{Higgs masses: Scenario A}

As already mentioned, in scenario A, from Eq.~(\ref{Hmasseig}) we get $\tilde{H}=H_2$, and we set $H_2$ to be the SM-like Higgs boson with mass $\sim 125~$GeV. 
The alignment limit, given in Ref.\cite{Das:2019yad}, can be seen explicitly in our case from Eq.~(\ref{Hmasseig}) and corresponds to 
\begin{eqnarray}
\sin(\alpha-\theta)=1 &\text{and} & \cos(\alpha-\theta)=0.
\label{alig:scenA}
\end{eqnarray}
 In this scenario, $H_2$ couples maximally to the gauge bosons, and $H_1$ is decoupled from the gauge bosons. The third neutral scalar, $h_0$, is always   decoupled from the gauge bosons due to the ${\cal Z}_2$ symmetry. A study of the masses in this scenario has been performed in \cite{Das:2014fea}, but with slightly different considerations as the ones taken here, as we explain below.

\subsubsection{Higgs boson: Scenario B }
In scenario B, we take $H_1$ as the SM-like Higgs boson, coupled maximally to the gauge bosons (here from Eq.~(\ref{Hmasseig}) we have $\tilde{H}=H_1$).   
The alignment limit in this scenario is expressed as
\begin{eqnarray}
\sin(\alpha-\theta)=0 &\text{and} & \cos(\alpha-\theta)=1.
\label{alig:scenB}
\end{eqnarray}
Although $H_2$ is always lighter than $H_1$,  as can be seen from the expressions for the masses Eqs.(\ref{H1masstree}) and (\ref{H2masstree}), it does not couple to the gauge sector in this scenario, thus it could escape experimental detection. This scenario has not been analysed  in the S3-3H model before. This  could be interesting in the context of a possible Higgs decay of an exotic scalar with mass $m_\Phi = 96$ GeV, as reported by CMS \cite{CMS-PAS-HIG-14-037} and discussed later.

Applying these alignment limits to the Higgs neutral masses $H_1$ and $H_2$,  Eqs. (\ref{H1masstree}) and (\ref{H2masstree}), the reduced expressions in each scenario can be obtained. 

\subsection{Numerical analysis and results}\label{Numerical}

 From the tree level Higgs mass expressions Eqs.~(\ref{c3e44}-\ref{c3e47}), (\ref{H1masstree}), and (\ref{H2masstree}), we can calculate the masses in terms of the Higgs self-couplings ($a,...,h$) and $\theta$, where $\tan\theta=2v_2/v_3$. 
We perform a scan on the eight self-couplings and $\tan\theta$ (see Eqs. (\ref{c3e5}) and (\ref{angtheta})). We produce $\mathcal{O}(10^{11})$   points with a pseudo-random generator, on these we first apply the  stability and unitarity constrains as given in Ref.~\cite{Das:2014fea}, to calculate the masses, and  then we take out all points where the  charged Higgs scalar masses are below $80$ GeV \cite{Agashe:2014kda, Zyla:2020zbs}. On the surviving points, we apply the alignment constraints in both A and B scenarios.  Finally, we impose a restriction on the mass of the respective SM-like Higgs boson.  This gives us the mass range at tree level for the scalars in this model, with the above restrictions.

A similar analysis on scenario A has been performed in \cite{Das:2014fea}, but in their analysis they restricted the mass of $h_0$ to be always heavier than  $m_{H_2}$ (the SM-like Higgs). Another difference is that we have  applied the alignment limit within an approximation, to allow for the possibility of a minimal coupling to the non-SM Higgs, and we also allowed for a range of masses for the SM-like Higgs boson. On the other hand, scenario B has not been analysed before.

\begin{figure}[htbp]
  \centering
\includegraphics[width=0.48\textwidth]{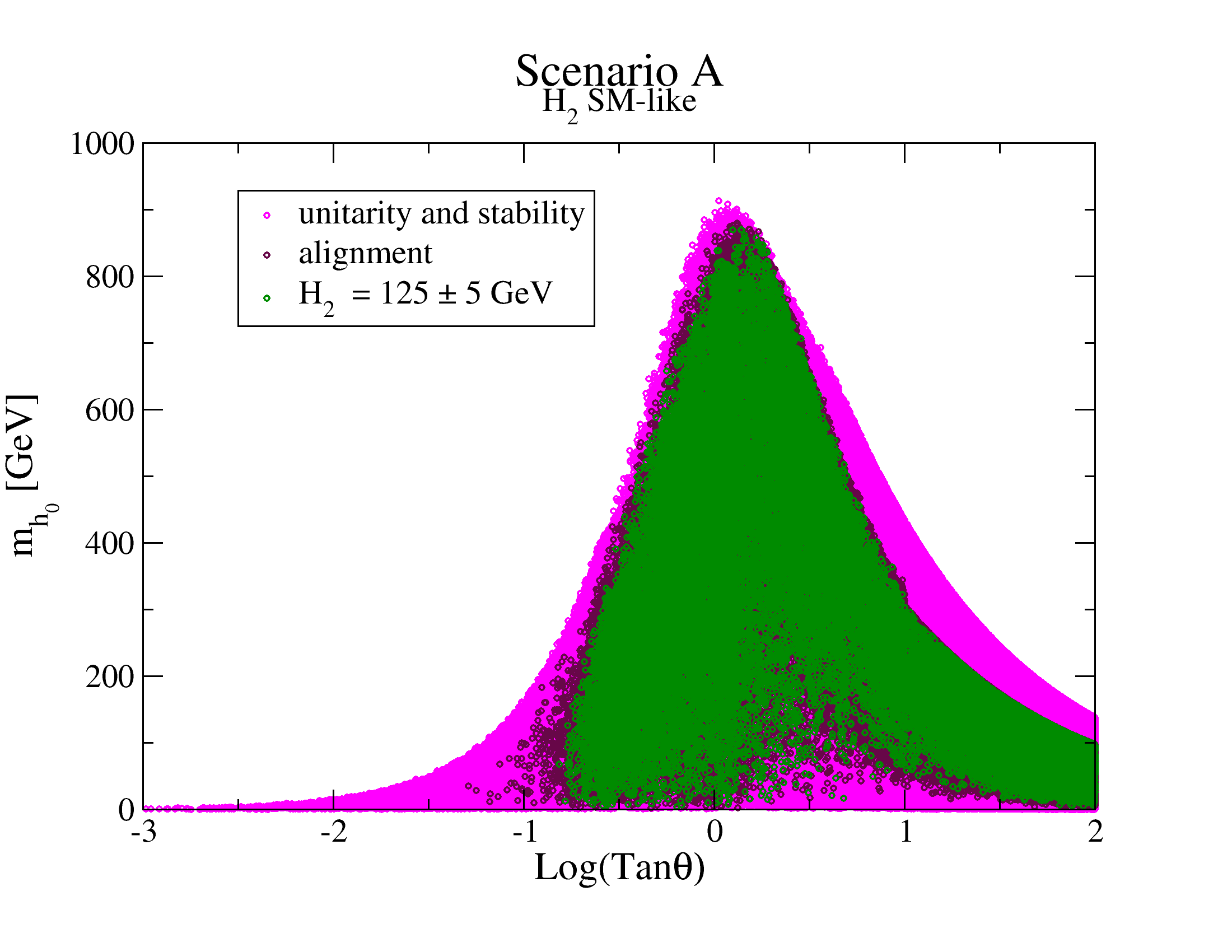}
\includegraphics[width=0.48\textwidth]{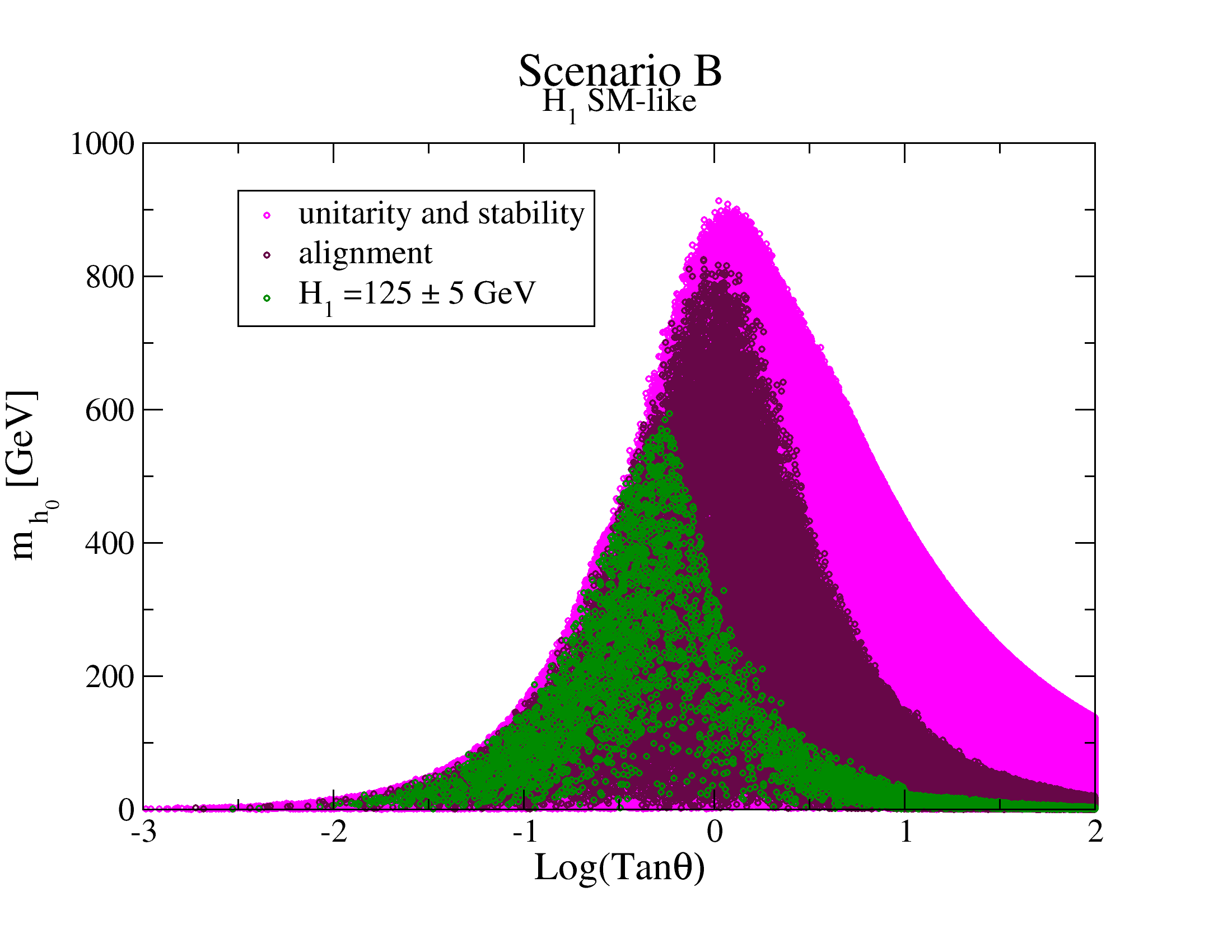}\\
\includegraphics[width=0.48\textwidth]{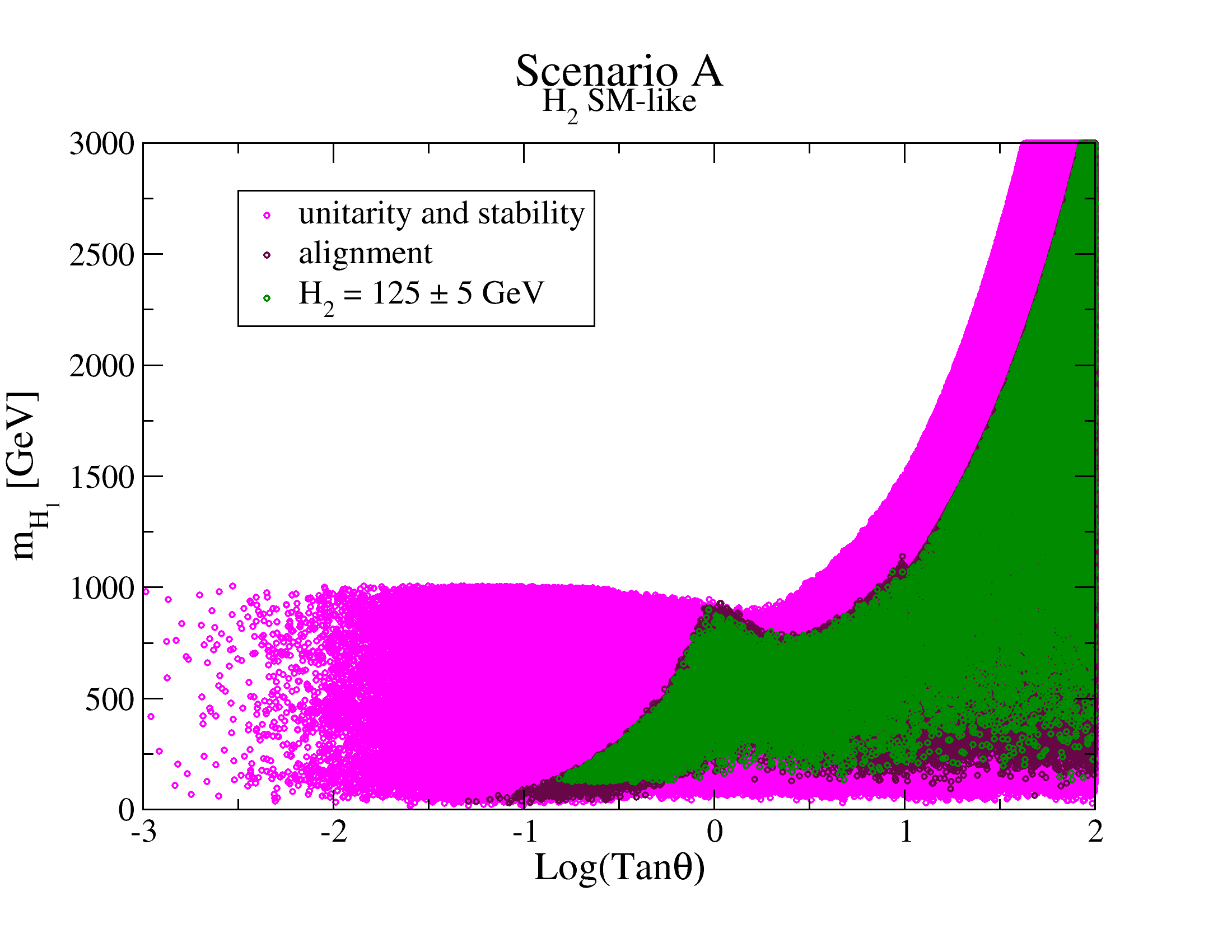}
\includegraphics[width=0.48\textwidth]{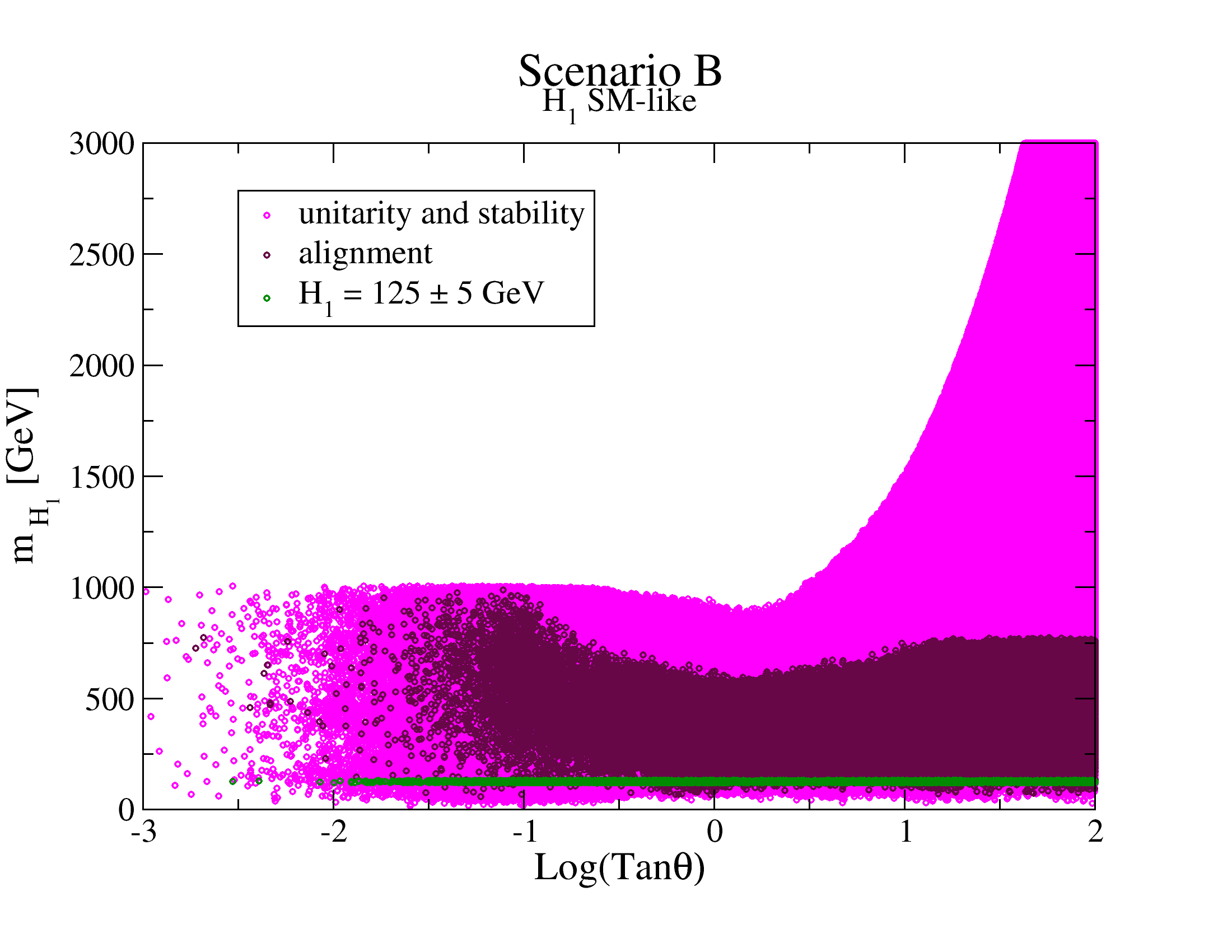}\\
\includegraphics[width=0.48\textwidth]{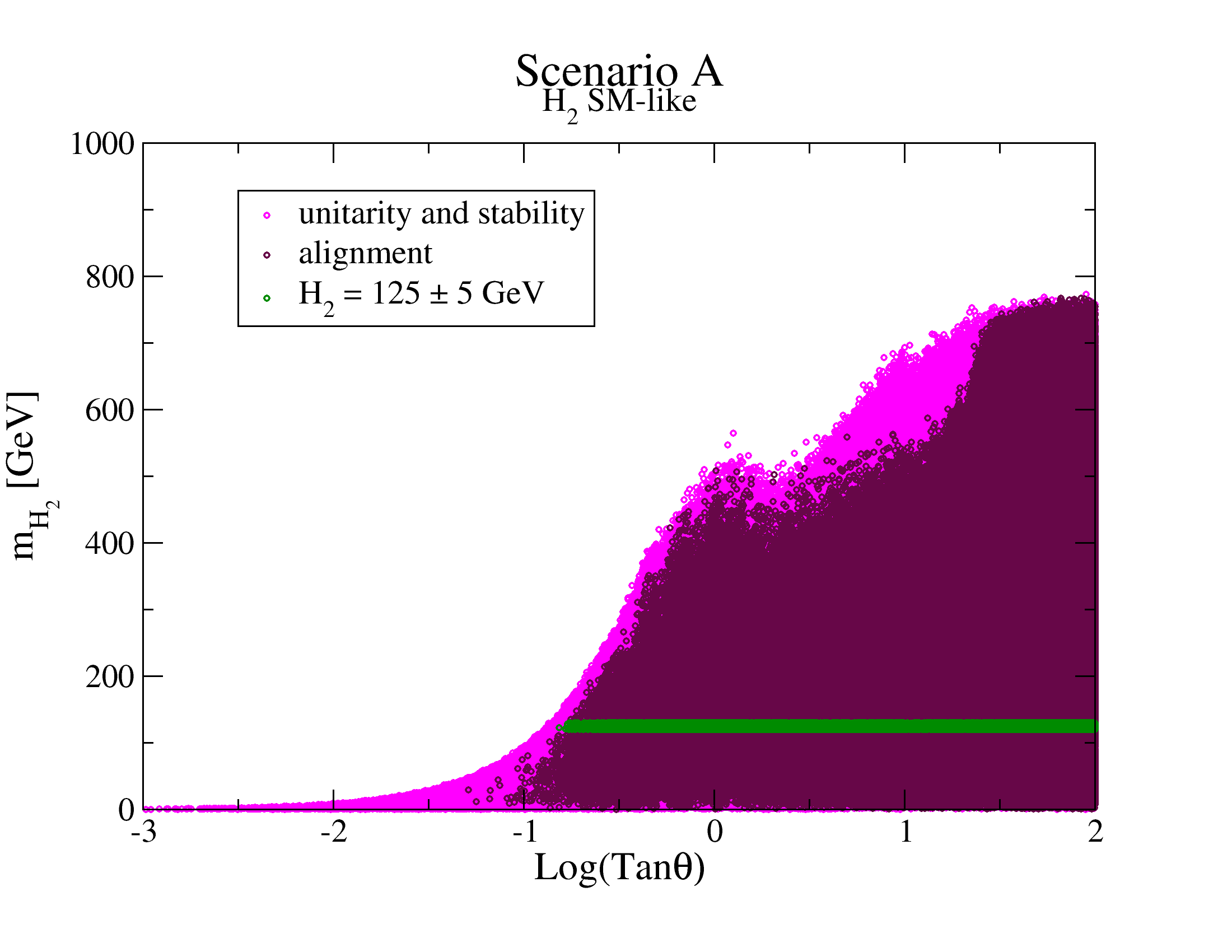}
\includegraphics[width=0.48\textwidth]{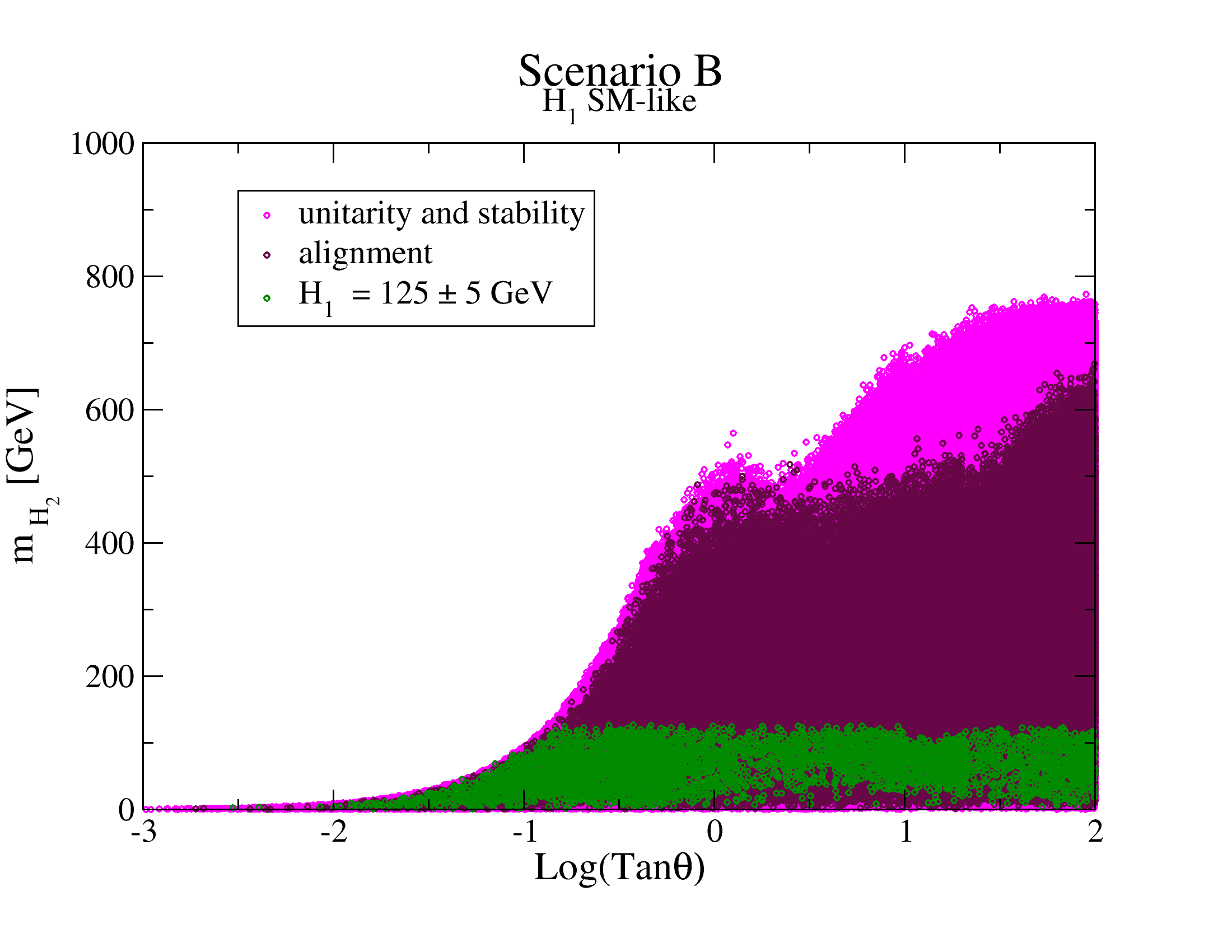}
\caption{ Dependence of the neutral scalar masses, $m_{h_0}$ and $m_{H_{1,2}}$,  on $\tan\theta$ for scenario A (left) and B (right). The {magenta points} comply with the unitarity and stability conditions, the {maroon points} comply further with the alignment conditions in each scenario.  Finally, the green ones have the SM-like mass restricted to  $m_{H_{2,1}} = 125 \pm 5$ GeV, respectively.
}
\label{figureAB-SM}
\end{figure}

In Figure \ref{figureAB-SM} we show the dependence of the three neutral scalars masses $m_{h_0}, m_{H_1}$, and $m_{H_2}$  on $\tan\theta$, 
for both scenarios, A in the left panel and B in the right one.  The upper two graphs correspond to the mass of $h_0$, the two graphs in the middle correspond to the mass of $H_1$ and the two bottom graphs correspond to the mass of the $H_2$. The magenta points correspond to the unitarity and stability constraints only (also excluding charged Higgs boson masses below 80 GeV), the maroon points, are a subset of the magenta ones, which also satisfy the respective alignment limit in each scenario, with a $10\%$ uncertainty  on the ($\alpha - \theta$) values, i.e., $\pm 0.1$. Finally, the green points are a subset of the maroon ones, in which the SM-like Higgs boson mass has been restricted to the 120-130 GeV range.  

As can be seen from the green points in Figure \ref{figureAB-SM}, the restriction of the $H_1$ mass to the 120-130  GeV range constrains the allowed values for $h_0$ in scenario B, much more strongly than the equivalent restriction in scenario A. In the case of $h_0$ the allowed upper bound for $m_{h_0} \sim 600$ GeV in scenario B, is lower than in scenario A, where $m_{h_0} \lesssim 900$ GeV. It can also be seen from  Figure~\ref{figureAB-SM}, that the Higgs neutral bosons could be degenerate  in mass, nevertheless once we restrict to the SM value for one of them, this possibility gets drastically reduced among $H_1$ and $H_2$,  although $h_0$ could be still degenerate in mass with the other two (at tree level). 

Notice that in both scenarios there exists the prospect of a lighter neutral Higgs boson to explain the possible  decay of a scalar with $m_\phi \sim 96$ GeV reported by CMS \cite{CMS-PAS-HIG-14-037,CMS-PAS-HIG-17-013}.  This was reported as a $\gamma  \gamma$ excess signal that could be due to a lighter neutral Higgs boson decay via a fermionic loop. This exotic Higgs boson role could be played by the lighter $H_2$ in scenario B, since it is always lighter than the SM Higgs, or by $h_0$ in both scenarios if it has  couplings to  fermions. This possibility, of a second lighter Higgs scalar consistent with this signal,  has been explored in SUSY models in \cite{Heinemeyer:2018wzl}. There are also  recent analyses along these lines in 2HDM and N2HDM   \cite{Haisch:2017gql,Biekotter:2019kde,Biekotter:2019mib}. Experimental bounds on possible decays of this type of Higgs bosons  will constrain further the parameter space.   

  \begin{figure}[htbp]
  \centering
\includegraphics[width=0.48\textwidth]{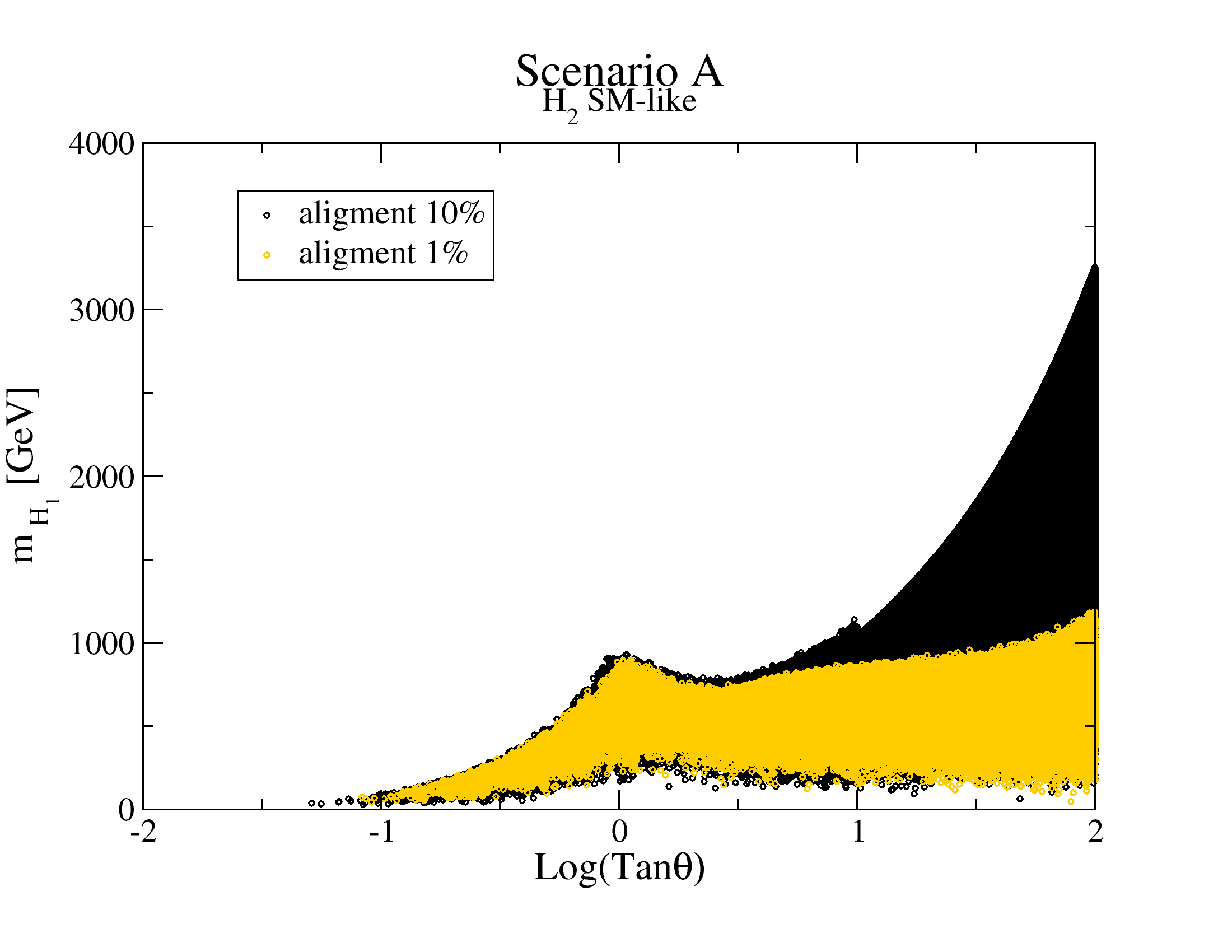}
\includegraphics[width=0.48\textwidth]{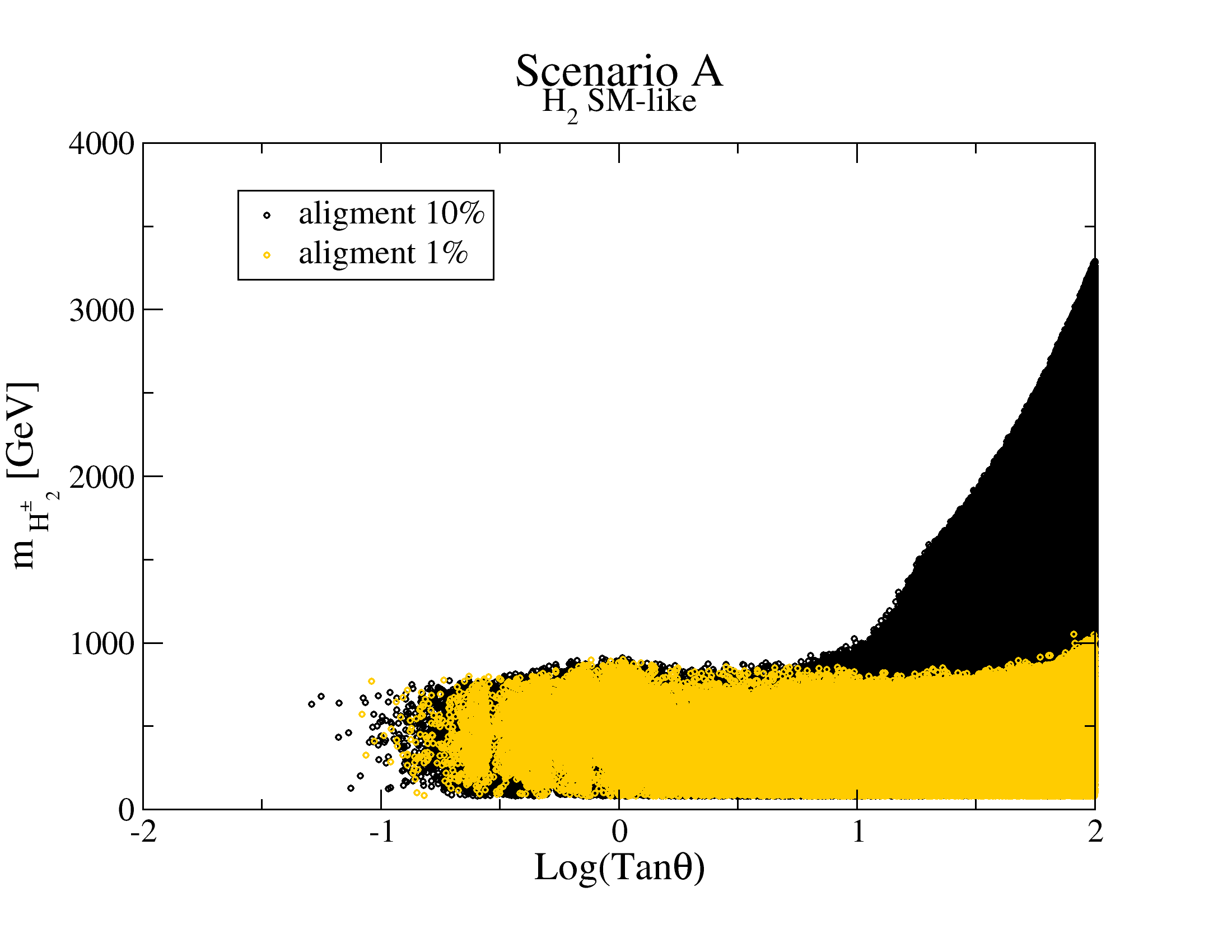}\\
\includegraphics[width=0.48\textwidth]{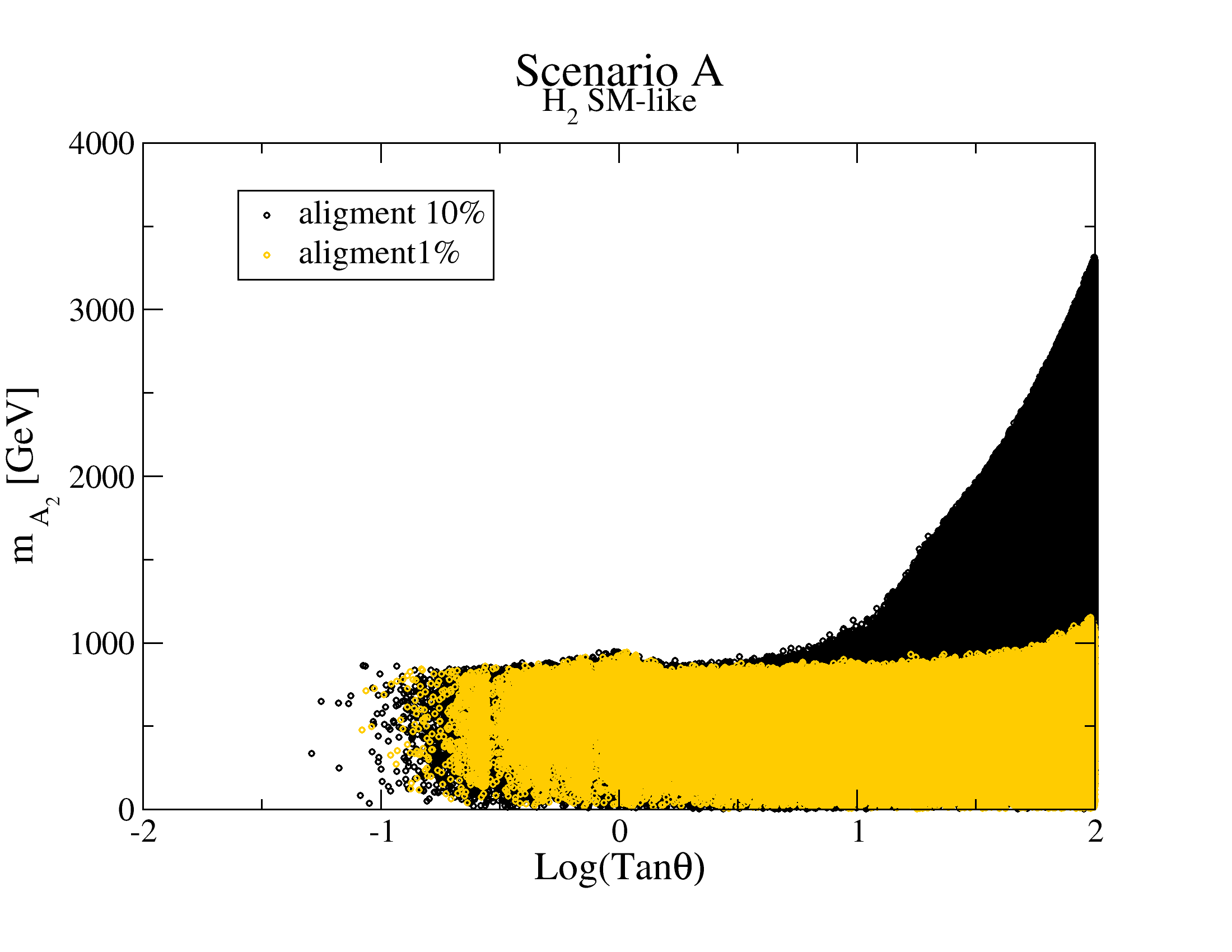}
\caption{ Dependence of the  masses $m_{H_1}$, $m_{H^\pm_2}$, and  $m_{A_2}$ on $\tan \theta$ for scenario A, applied with a $10\%$ uncertainty (black points) and $1\%$ uncertainty (yellow points) on $(\alpha-\theta)$. The points shown comply with the unitarity and stability conditions, and the restriction of $m_{H_2} = 125 \pm 5$ GeV.}
\label{figure10to1}
\end{figure}

In Figure~\ref{figure10to1} we present the masses of $H_1$, $H^\pm_2$ and $A_2$ in scenario A which satisfy the alignment limit, Eqs.~(\ref{alig:scenA}),  applied  with a $10\%$ and  $1\%$ uncertainty on the $(\alpha -\theta)$ values. In the graphs we show only the masses which are affected by the precision of the values in $(\alpha -\theta)$. The black points are within  $10\%$ of the alignment limit and the yellow ones within $1\%$. The restriction to an alignment limit with  $1\%$ precision only  appears as a noticeable difference  for values of
$\log(\tan\theta)>1$,  where the values of the masses are constrained to be below $\sim 1$ TeV. The rest of the masses in scenario A and the masses in  scenario B are affected only very slightly by changing the precision in the alignment limit.  

 \begin{figure}[htbp]
  \centering
\includegraphics[width=0.48\textwidth]{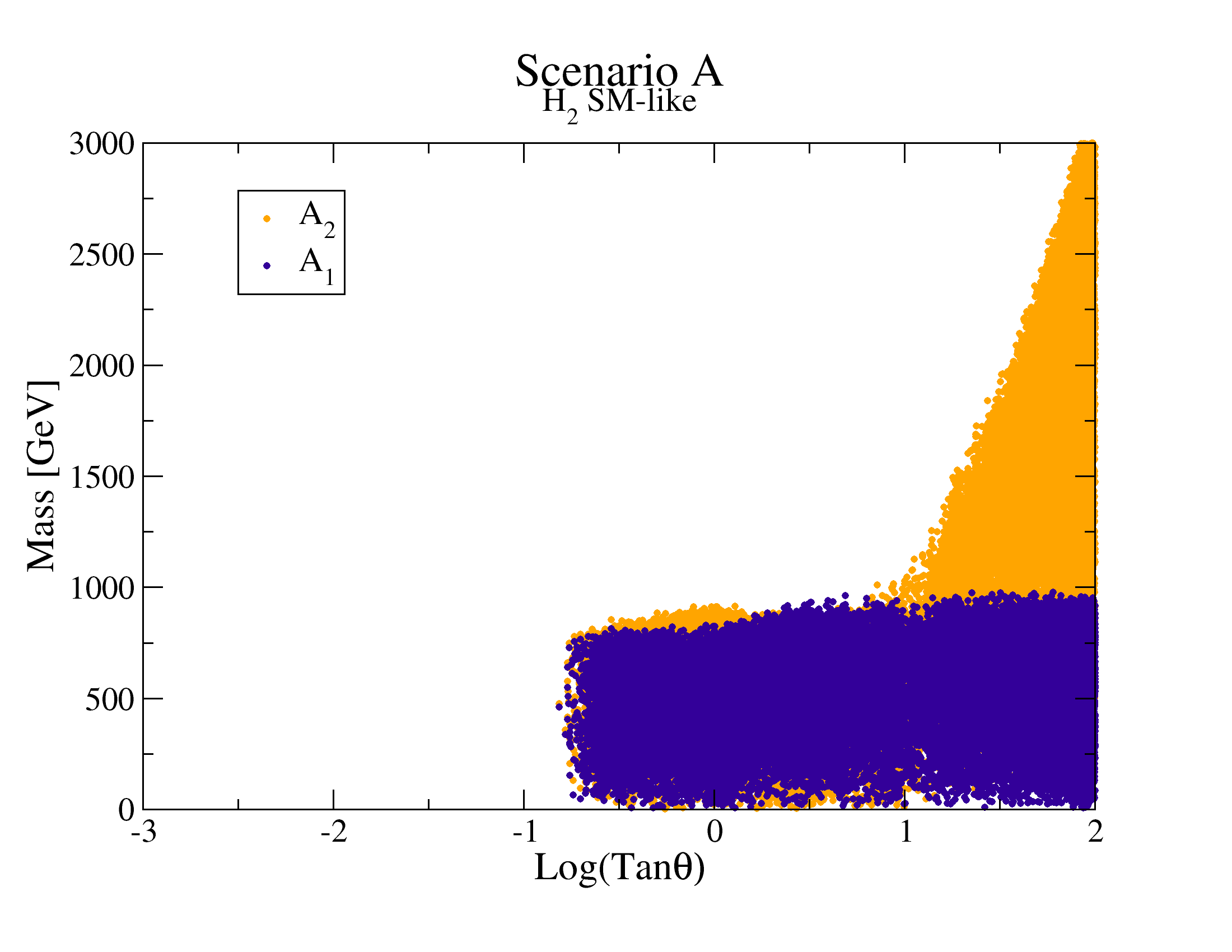}
\includegraphics[width=0.48\textwidth]{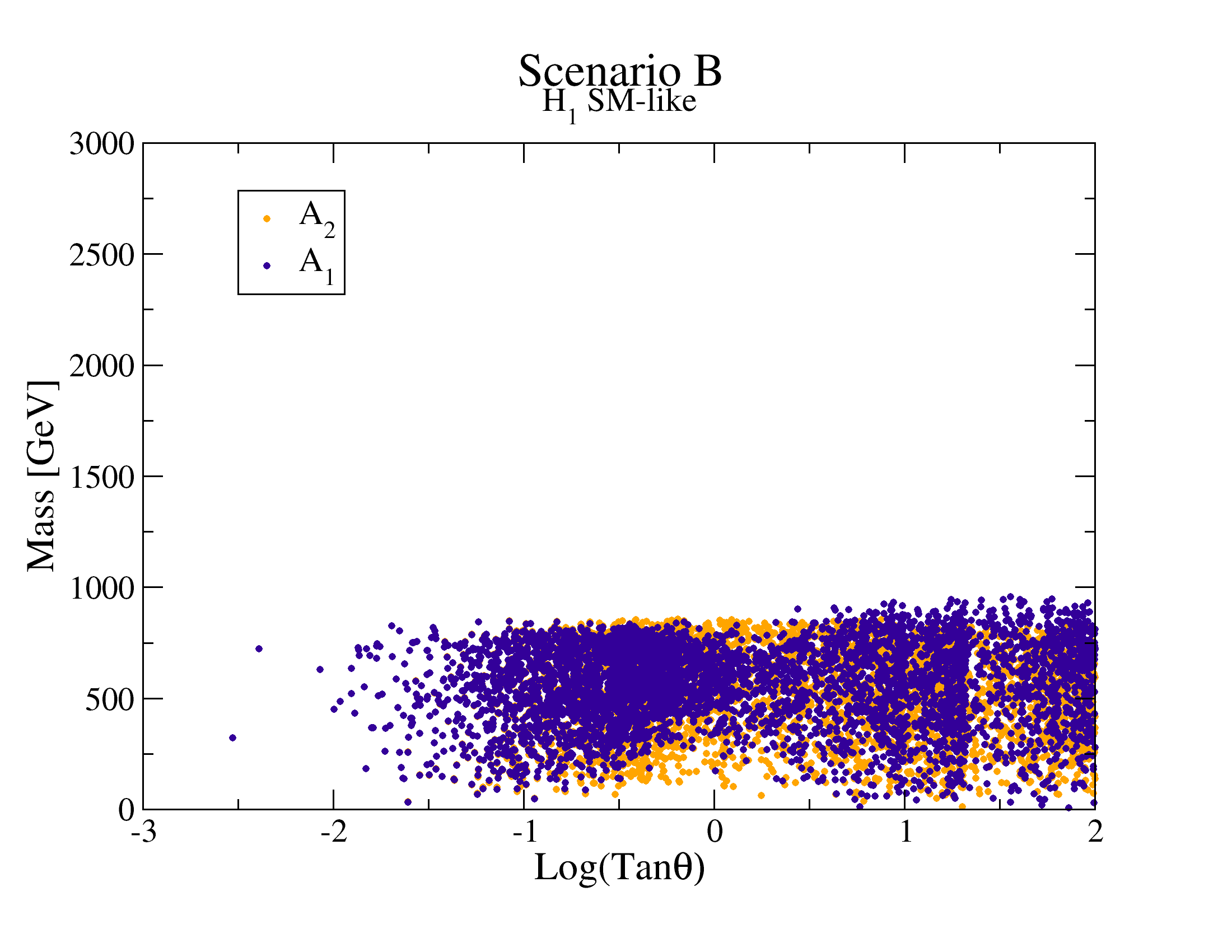}\\
\includegraphics[width=0.48\textwidth]{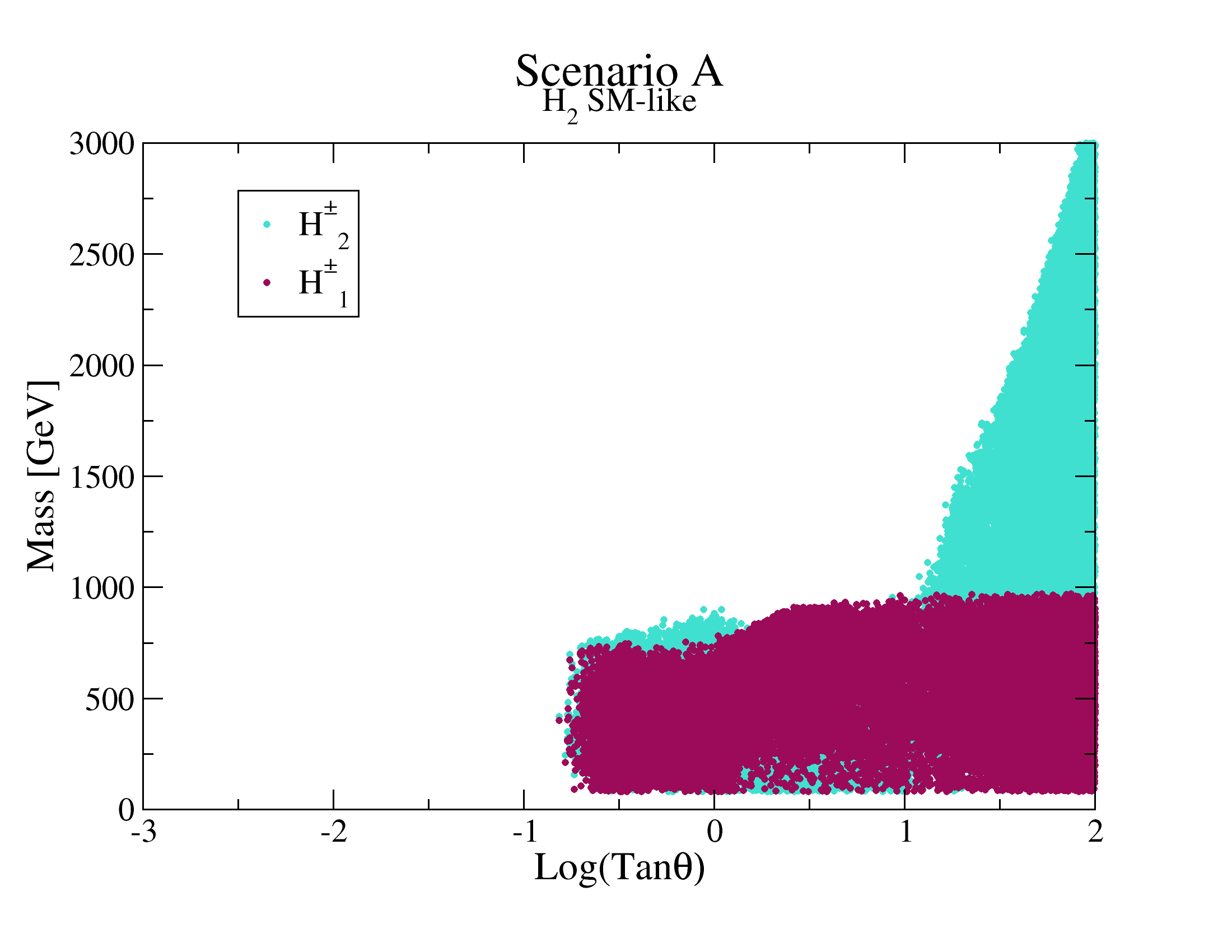}
\includegraphics[width=0.48\textwidth]{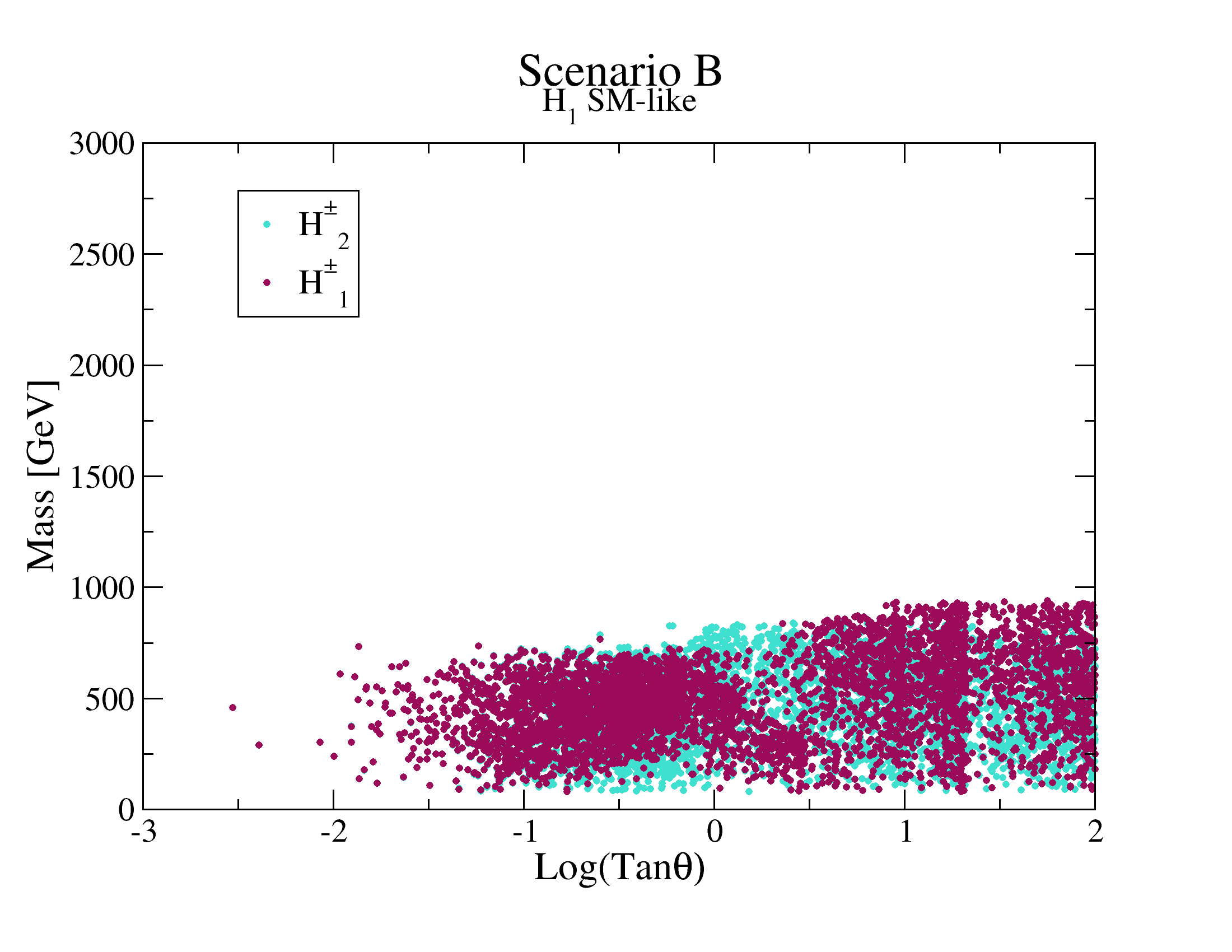}
\caption{
Dependence of the two pseudoscalar masses $m_{A_{1,2}}$ (upper panel) and charged scalars $m_{H_{1,2}^{\pm}}$ (lower panel) on $\tan\theta$. The points shown comply with the constraints of previous figures plus the bounds on the SM-like Higgs boson mass for each scenario. }
\label{figureSeudoCharge}
\end{figure}

Finally in Figure~\ref{figureSeudoCharge}, we show the pseudoscalar  masses $m_{A_{1,2}}$, and charged Higgs masses $m_{H_{1,2}^{\pm}}$ dependence on $\tan\theta$, after all constraints have been applied, including when one of the neutral scalars is restricted to be the SM-like Higgs boson.  As already mentioned, points where $m_{H_{1,2}^{\pm}} < 80$ GeV have already been excluded in every figure. The figures are shown with a precision of 10\% in the alignment limits. The first two graphs  correspond to the masses of the pseudoscalars in each scenario (A in the left, B in the right), the orange points represent the mass of $A_2$ and the purple ones represent the mass of $A_1$. The last two graphs  correspond to the masses of the charged scalars in each scenario, the cyan points represent the mass of $H_2^\pm$ and the pink ones represent the mass of $H_1^\pm$. From this figure we can see what are the upper bounds for these masses, and  that for small values of $\tan\theta$ they will be constrained to be below $\sim 1$ TeV. It can also be seen that there are regions in parameter space where the masses can be very close in value.  This is relevant when calculating the  values of the trilinear and quartic couplings, as well as the possible contributions to the oblique parameters, as will be discussed in next section.

We want to highlight here that these are tree level masses, 
 radiative corrections will change their actual theoretical value, which in the case of the one identified as the SM Higgs,  we have taken into account with a conservative uncertainty of $\pm 5$ GeV.
A next-to-leading order (NLO) calculation of the masses should be done in order to give more accurate theoretical predictions, that could be tested at the LHC or future colliders. Work in this  direction is proposed in \cite{Chakrabarty:2018yoy}. NLO analytical expressions for the scalar contributions are given in \ref{loopHiggs-masses}. In order to perform a numerical calculation for these loop corrections, we need to establish the parameter dependence for the trilinear and quartic Higgs couplings, which we have calculated and whose expressions are given in section \ref{scalars} and in Appendix \ref{appendixA}.

An analysis of the scalar sector of a similar model with 4 Higgs doublets (S3-4H) \cite{Espinoza:2018itz}, where the fourth doublet is inert and  with some considerations in the Yukawa sector, shows that the region that satisfies the bounds on extra scalar searches \cite{Bechtle:2015pma}, prefers values  of $\tan\theta \lesssim 5$.  As we will see in next section, this might also apply here.

\section{Higgs couplings}
\label{gcouplings}

In this section we calculate the trilinear and quartic couplings among the Higgs bosons, as well as with the gauge bosons, and  give the analytical expressions in terms of the physical parameters. We analyse the contributions of these couplings to the neutral one-loop scalar mass matrix. We give the explicit expressions for scenario A. 

\subsection{Gauge-Higgs couplings}
\label{G-Hcoupl}

We expand the scalar kinetic Lagrangian term, Eq. (\ref{kinHiggs}), to calculate the couplings to the EW gauge bosons, performing the usual  EW rotation on the gauge fields $W_\mu^3$ and $B_\mu$.

The residual ${\cal Z}_2$ symmetry manifests itself also in the gauge-Higgs couplings.  We show here the couplings of gauge bosons with the three neutral scalars, the rest of the couplings are given in  appendix \ref{appendixA},  in this expression we have not taken into account the combinatorial factor from two identical particles in the Lagrangian term. Notice that  $h_0$ does not couple in a single scalar coupling with the gauge bosons but it does in pairs with gauge bosons:  

\begin{eqnarray}
g_{h_0W^\pm W^\mp} =  0, \;\;\;
g_{h_0Z Z} =  0;
\label{h0VV}
\end{eqnarray}

\begin{eqnarray}
g_{H_1W^\pm W^\mp} = \frac{2 M^2_W \cos(\alpha - \theta)g^{\mu \nu}}{v}, \;\;\; 
g_{H_2W^\pm W^\mp} = \frac{2 M^2_W \sin(\alpha - \theta)g^{\mu \nu} }{v};
\label{H12WW}
\end{eqnarray}

\begin{eqnarray}
g_{H_1ZZ}= \frac{M^2_Z \cos(\alpha-\theta)g^{\mu\nu}}{v}, \;\;\;
g_{H_2ZZ}= \frac{M^2_Z \sin(\alpha-\theta)g^{\mu\nu}}{v};
\label{H12ZZ}
\end{eqnarray}

\begin{eqnarray}
g_{h_0h_0W^\pm W^\mp} = \frac{ M^2_W g^{\mu \nu}}{v^2},\;\;\;
g_{h_0h_0ZZ}= \frac{M^2_Z g^{\mu\nu}}{2v^2};
\label{h0h0VV}
\end{eqnarray}

\begin{eqnarray}
g_{H_1H_1W^\pm W^\mp} = \frac{ M^2_W g^{\mu \nu}}{v^2}, \;\;\; 
g_{H_2H_2W^\pm W^\mp} = \frac{ M^2_W g^{\mu \nu}}{v^2} ;
\end{eqnarray}

\begin{eqnarray}
g_{H_1H_1ZZ}= \frac{M^2_Z g^{\mu\nu}}{2v^2},\;\;\;
g_{H_2H_2ZZ}= \frac{M^2_Z g^{\mu\nu}}{2v^2}.
\end{eqnarray}

The form of the cubic couplings reflects the residual ${\cal Z}_2$ symmetry.  The couplings of the gauge bosons with $h_0$ vanish, as expected.  The expressions for the gauge couplings to the other two neutral scalars $H_{1,2}$ are similar to the ones in the 2HDM \cite{Gunion:2002zf}, reflecting the fact that  these two scalars decouple from $h_0$ due to the ${\mathcal Z}_2$ symmetry. 

For the trilinear couplings, in the exact alignment limit, only the  ones corresponding to the SM-like Higgs boson in each scenario will be different from zero.

\subsection{Higgs-Higgs couplings}
\label{scalars}
The trilinear and quartic Higgs couplings will be important to estimate radiative corrections, in particular for the SM Higgs, as well as possible loop  contributions to physical processes. Previously, the trilinear couplings for the neutral scalars in the  3HDM with $S_3$ symmetry  were reported in \cite{Barradas-Guevara:2014yoa}, nevertheless our results differ from the ones calculated there. On the other hand, our expressions for the trilinear couplings do coincide with the presence of a residual ${\cal Z}_2$,  as reported in \cite{Das:2014fea}. Besides the confirmation of this residual symmetry, we additionally show that the couplings reduce to the SM ones for the particular alignment limits.

The  self-couplings given in the Higgs potential, Eq. (\ref{Hpotential}) can be obtained in terms of phy\-si\-cal parameters  using the rotation matrices. 
The  angle $\alpha$ given in Eq.~(\ref{scalargamma}) 
can be re-written using the relations we obtained in Eq.~(\ref{H12masses}). Thus, we can write Eq.~(\ref{MelemSPh}) in terms of the physical Higgs masses and rotation angle $\alpha$.
Moreover, also using Eqs.~(\ref{c3e44})-(\ref{c3e47}) and Eqs.(\ref{h0masstree})-(\ref{H12masses}), we obtain expressions for the self-couplings in the scalar potential,  Eq.~(\ref{Hpotential}), given in terms of the physical parameters {\it i.e.} masses, {\it vevs}, and rotation angles ($v, m^2_{h_0},m^2_{H_1},m^2_{H_2},m^2_{A_1},m^2_{A_2},m^2_{H_1^{\pm}},m^2_{H_2^{\pm}},\tan\alpha, \tan\theta$), as
 \begin{eqnarray}
 a&=& \frac{1}{v^2\cos^2\theta}\left[m^2_{H_1}\cos^2\alpha + m^2_{H_2}\sin^2\alpha - \frac{1}{9}m^2_{h_0}\tan^2\theta \right],\\ \label{c2e39}
 b&=& \frac{1}{v^2}\left[  \frac{\sin2\alpha}{\sin2\theta}(m^2_{H_1} - m^2_{H_2}) + \frac{m^2_{h_0}}{9\cos^2\theta} + 2m^2_{H^\pm_2} \right],\\ \label{c2e40}
 c&=& \frac{1}{v^2\sin^2\theta}\left[m^2_{H_1}\sin^2\alpha + m^2_{H_2}\cos^2\alpha - \frac{1}{9}m^2_{h_0} - m^2_{H^\pm_2}\cos^2\theta + m^2_{H^\pm_1} \right],\\ \label{c2e41}
d&=&\frac{1}{v^2\sin^2\theta}\left[(m^2_{H^\pm_1} - m^2_{A_1})- (m^2_{H^\pm_2} - m^2_{A_2})\cos^2\theta \right],\\
e&=&	-\frac{4m_{h_0}^2}{9v^2\sin2\theta},\\
f&=& \frac{1}{v^2}\left[ \frac{m^2_{h_0}}{9\cos^2\theta} +m^2_{A_2} - 2m^2_{H^\pm_2} \right],\\
g&=&\frac{1}{v^2\sin^2\theta}\left[\frac{4}{9}m^2_{h_0} + m^2_{H^\pm_2}\cos^2\theta - m^2_{H^\pm_1} \right],\\
h&=&\frac{1}{v^2}\left[\frac{m^2_{h_0}}{9\cos^2\theta} - m^2_{A_2}\right].\label{c2e70}
\end{eqnarray}
This parameterization of the scalar potential self-couplings differs slightly from the ones presented in other works, as in \cite{Barradas-Guevara:2014yoa,Das:2014fea}, due to our normalization of the couplings in the scalar potential. 

From the scalar potential we can get the trilinear scalar couplings as usual, where all possible combinations are given from the terms considered in the  potential. 
 \begin{eqnarray}
 -i\lambda_{ijk} = \frac{-i\partial^3 V}{\partial H_i \partial H_j \partial H_k}. \label{HHH}
 \end{eqnarray}

 As we already mentioned, the ${\cal Z}_2$ residual symmetry present will imply a vanishing trilinear $h_0h_0h_0$ due to  its odd charge under ${\cal Z}_2$, we explicitly confirmed this and also obtain the other trilinear and quartic scalar couplings. These couplings are essential in order to determine experimentally the shape of the actual Higgs potential. To this end,  a one-loop calculation of the self-energy corrections and vertices should be performed.  Moreover, it is possible to restrict parameters from the SM Higgs boson mass corrections, as we are going to consider in next section. 
 
The following analytical expressions are the scalar-scalar couplings written in the physical basis and in terms of the physical parameters: \footnote{As we mentioned in the previous section the symmetry factor n! has to be added in front of the couplings for n identical particles in the vertex.}

\begin{eqnarray}
g_{h_0h_0h_0} = 0 , \label{3h0}
\end{eqnarray}

\begin{eqnarray} g_{H_2H_2H_2}& = & -\frac{1}{v \, s_{2\theta}}\left[  m^2_{h_0} \frac{ c^3_{\alpha -\theta}}{ 9 c^2_\theta} + m^2_{H_2} \left( c^2_\alpha c_{\alpha - \theta} - s_{ \alpha} s_{\theta} \right) \right], \label{c2e43}
\end{eqnarray}

\begin{eqnarray} g_{H_1H_1H_1}& = & \frac{1}{v\,s_{2\theta}}\left[  m^2_{h_0} \frac{s^3_{\alpha-\theta} }{ 9c^2_\theta } - m^2_{H_1}\left( c^2_\alpha s_{\alpha - \theta} - s_{ \alpha} c_{\theta} \right) \right] ,
\end{eqnarray}

\begin{eqnarray}
g_{h_0h_0H_1}&=& \frac{1}{v\,s_{2\theta}}(m^2_{h_0}s_{\alpha + \theta} +m^2_{H_1} s_\alpha c_\theta  ) ,\label{h0h0H1}
\end{eqnarray} 

\begin{eqnarray}
g_{h_0h_0H_2} &=&  -\frac{1}{v\,s_{2\theta}}(m^2_{h_0}c_{\alpha + \theta} + m^2_{H_2}c_\alpha c_\theta  ) , \label{h0h0H2}
\end{eqnarray}

\begin{eqnarray}
g_{H_1H_1H_2} &=&- \frac{s_{\alpha - \theta}}{vs_{2\theta}} \Bigg( m^2_{h_0}\left(\frac{s_{2(\alpha - \theta)} }{6c^2_\theta}  \right) +m^2_{H_1}s_{2\alpha}  + \frac{ m^2_{H_2} s_{2\alpha}}{2} \Bigg) , \label{c2e47}
\end{eqnarray}

\begin{eqnarray}
g_{H_1H_2H_2} &=& \frac{c_{\alpha-\theta}}{vs_{2\theta}} \Bigg( m^2_{h_0}\left(\frac{s_{2(\alpha - \theta)} }{6c^2_\theta}  \right) +\frac{m^2_{H_1} s_{2\alpha}}{2}  +  m^2_{H_2}s_{2\alpha}   \Bigg) ,\label{c2e48}
\end{eqnarray}
here we use the reduced notation  $s_{x}\equiv \sin x$, $c_{x}\equiv \cos x$ and $t_{x}\equiv \tan x$.

In the following we show the analytical expressions for the  trilinear couplings between scalars and pseudoscalars, as well as with the Goldstone boson. The residual $\mathcal{Z}_2$ symmetry is also evident in the allowed couplings with the pseudoscalars (the forbidden ones are not present), which are given as:\footnote{The couplings with the Goldstone boson may be important depending on the renormalization procedure used.}

\begin{eqnarray}
g_{A_1A_1H_1}= \frac{1}{vs_\theta}(-m^2_{h_0}\frac{s_{\alpha-\theta}}{6c_\theta} + \frac{1}{2} m^2_{H_1}s_\alpha + m^2_{A_1}s_\alpha - m^2_{A_2}c_\theta s_{\alpha-\theta} ),
\end{eqnarray}

\begin{eqnarray}
g_{A_1A_1H_2}= \frac{1}{vs_\theta}(\frac{m^2_{h_0}c_{\alpha-\theta}}{6c_\theta} - \frac{m^2_{H_2}c_\alpha}{2} -m^2_{A_1}c_\alpha + m^2_{A_2}c_\theta c_{\alpha-\theta} ),
\end{eqnarray}

\begin{eqnarray}
g_{A_2A_2H_1} = \frac{1}{v s_{2\theta}}\left(\frac{m^2_{h_0}s_{\alpha-\theta}}{9 c^2_\theta} +  m^2_{H_1}\left( s_\alpha c^3_\theta-c_\alpha s^3_{\theta} \right) + m^2_{A_2}s_{2\theta}c_{\alpha-\theta} \right),
\end{eqnarray}

\begin{eqnarray}
g_{A_2A_2H_2}= \frac{1}{v s_{2\theta}}\left(-\frac{m^2_{h_0}c_{\alpha-\theta}}{9 c^2_\theta} + m^2_{H_2}\left(s_\alpha s^3_\theta- c_\alpha c^3_\theta  \right) + m^2_{A_2}s_{2\theta}s_{\alpha-\theta} \right),
\end{eqnarray}

\begin{eqnarray}
g_{A_1A_2h_0}= \frac{2}{3vs_{2\theta}} \left( -m^2_{h_0} (c_{2\theta} + c^2_\theta) + 3m^2_{A_1}c^2_\theta - 3m^2_{A_2}c^2_\theta  \right), \label{A1A2h0}
\end{eqnarray}

\begin{eqnarray}
g_{H_2G_0G_0 }= \frac{m^2_{H_2}s_{\alpha - \theta}}{2v},\;\;\;
g_{H_1G_0G_0} = \frac{m^2_{H_1}c_{\alpha - \theta}}{2v},
\end{eqnarray}
\begin{eqnarray}
g_{H_2G_0A_2}= \frac{c_{\alpha-\theta}}{v}(m^2_{H_2} - m^2_{A_2}),\;\;\;
g_{H_1G_0A_2}= \frac{s_{\alpha-\theta}}{v}(-m^2_{H_1} + m^2_{A_2}),
\end{eqnarray}

\begin{eqnarray}
g_{h_0G_0A_1}= \frac{1}{v}(m^2_{h_0} - m^2_{A_1} ).
\end{eqnarray}

The couplings involving charged Higgses can be found in the Appendix.
The diagonalization of the mass matrices leaves a structure which is similar to a 2HDM in the $2\times 2$ block, which is $\Zd$ even ($H_2, H_1, H^\pm_2,A_2$), plus  decoupled particles which are $\Zd$ odd ($H^\pm_1, A_1, h_0$). Notice though, that the allowed parameter spaces of both models might be different due to the extra $\Zd$ odd scalar particles and their associated couplings in our model as compared to the 2HDM.  
 For instance, we have possible extra channels for SM Higgs boson production and  di-Higgs production, which in the LHC could be $pp\to Hqq$ and $pp\to HH\to b\bar{b}\gamma \gamma, b\bar{b}b\bar{b}, b\bar{b}\tau^+\tau^-$ respectively, which will happen via the exchange of  odd $\Zd$ scalars, and  will depend on the scalar trilinear couplings and on the Yukawa couplings of the model.
There will also be loop corrections to the SM Higgs mass due to the $\Zd$ odd particles, which are absent  in the 2HDM. These  processes may provide a  possible way to differentiate this model from the 2HDM in the alignment limit \cite{Gunion:2002zf}. These type of analyses should be performed in order to have more precise predictions and bounds on the parameter space of the model.

From expressions (\ref{A1A2h0},\ref{h0Hc1Hc2}) it can be seen that the trilinear couplings among $h_0$ and the other $\Zd$ odd particles are allowed. As already mentioned,  $h_0$ may be a dark matter candidate, provided it has no couplings to SM fermions and it is the lightest particle in the ${\cal Z}_2$ odd sector, thus $m_{H^\pm_1},m_{A_1} > m_{h_0}$.

The quartic scalar couplings written in the physical basis and in terms of the physical parameters are calculated from
\begin{eqnarray}
 -i\lambda_{ijkl} = \frac{-i\partial^4 V}{\partial H_i \partial H_j \partial H_k\partial H_l}. \label{HHHH}
 \end{eqnarray}
We also give the analytical expressions for the quartic self-scalar couplings written in the physical basis and in terms of the physical parameters. We are interested in particular in the couplings with $H_{1,2}$, in order to compare with the SM ones in the alignment limits.
The quartic couplings also necessary to calculate the one-loop  corrections to the Higgs bosons masses. 
We give here some explicit four scalar couplings examples, the rest of the couplings can be found at the Appendix \ref{AppA2}:
\begin{eqnarray} g_{h_0h_0h_0h_0} &=& \frac{1}{24 v^2s^2_\theta} \Bigg(m^2_{h_0} + 3m^2_{H_1} s^2_\alpha + 3 m^2_{H_2}c^2_{\alpha} \Bigg)  , \label{c2e51-2}
\end{eqnarray}

\begin{eqnarray}
\notag g_{H_1H_1H_1H_1} &=& \frac{1}{2v^2s^2_{2\theta}} \Bigg( m^2_{h_0}s^3_{\alpha -\theta} \frac{(s_{\alpha -\theta} + 2s_{\alpha + \theta} )}{9c^2_\theta} \\
&& + m^2_{H_1}(s^2_\alpha s_{\alpha -\theta} +c_{\alpha}s_{\theta})^2  + m^2_{H_2}\frac{s^2_{2\alpha}s^2_{\alpha -\theta}}{4}  \Bigg) , \label{c2e50}
\end{eqnarray}	

\begin{eqnarray}
\notag g_{H_2H_2H_2H_2} &=& \frac{1}{2v^2s^2_{2\theta}} \Bigg( m^2_{h_0}c^3_{\alpha -\theta} \frac{(c_{\alpha -\theta} + 2c_{\alpha + \theta} )}{9c^2_\theta} \\
&& + m^2_{H_1}\frac{s^2_{2\alpha}c^2_{\alpha -\theta}}{4} + m^2_{H_2}(c^2_\alpha c_{\alpha -\theta} -s_{\alpha}s_{\theta})^2  \Bigg) . \label{c2e50-2}
\end{eqnarray}

\subsection{Couplings in scenario A}
\label{Coupl-alignA}
We show here how the scalar couplings are  reduced in the  alignment limit of scenario A. 
Recalling that the alignment limit is given as, $\sin(\alpha-\theta)= 1$, $\cos(\alpha-\theta)= 0$, the trigonometric functions for $\alpha$ and $\theta$ satisfy the following relations
\begin{eqnarray}
\sin{\alpha}= \cos{\theta};&
\cos{\alpha}=-\sin{\theta};& 
\sin2(\alpha-\theta)
= 0;\notag
\end{eqnarray}
\begin{eqnarray}
\cos(3\alpha-\theta)
=\,\sin2\theta; &
\sin(\alpha+\theta) = \, \cos2\theta; &
\cos(\alpha+\theta)
= \,-\sin2\theta.
\end{eqnarray}

In scenario A in the alignment limit,  the Higgs boson $H_2$ trilinear  coupling coincides exactly with the trilinear coupling of the SM Higgs boson $\lambda_{SM}$,
\begin{eqnarray} g_{H_2H_2H_2}& = & \frac{1}{v \, s_{2\theta}}\left[ m^2_{H_2} s_{ \alpha} s_{\theta}  \right]=\frac{1}{2v}\frac{ s_{ \alpha}}{ c_{\theta}}m^2_{H_2}=\frac{m^2_{H_2}}{2v}~ ~\equiv \lambda_{SM}.
\label{triH2align}
\end{eqnarray}
And the $H_1$ trilineal couplings reduces to
\begin{eqnarray} g_{H_1H_1H_1}& = & \frac{1}{v\,s_{2\theta}}\left[   \frac{1 }{ 9c^2_\theta }m^2_{h_0} - s^2_\theta m^2_{H_1}  \right]=\frac{1}{v\,s_{2\theta}c^2_\theta}\left[   \frac{1 }{ 9 }m^2_{h_0} - \frac{1}{2}s_{2\theta} m^2_{H_1}  \right].
\label{triH1align}
\end{eqnarray}
The $H_2$ quartic  coupling (\ref{c2e50}) also reduces exactly to the SM one in the alignment limit,
\begin{eqnarray}
g_{H_2H_2H_2H_2} &=& \frac{1}{2v^2s^2_{2\theta}}
m^2_{H_2}(-s^3_\theta c_{\theta} -c^3_{\theta}s_{\theta})^2
=\frac{m^2_{H_2}}{8v^2}
\label{H2H2H2-A}.
\end{eqnarray}
The $H_2-h_0$ quartic coupling reduces in this limit to
\begin{eqnarray}
g_{H_2H_2h_0h_0} &= &\frac{1}{v^2 s_{2\theta}} \Bigg(\frac{1}{6}m^2_{h_0}
3s_{2\theta} 
+\frac{1}{4} m^2_{H_2} s_{2\theta} \Bigg) =\frac{1}{4v^2}(2m^2_{h_0}+m^2_{H_2})~.
\label{DL-H2H2h0h0}
\end{eqnarray}

Some of the reduced scalar couplings for scenario A depend only on the masses involved, and are given as 
\begin{eqnarray}
g_{H_2h_0h_0}=\frac{1}{2v}(m^2_{H_2}+2m^2_{h_0}),& g_{H_2A_1A_1}=\frac{1}{2v}(m^2_{H_2}+2m^2_{A_1}), &
g_{H_2A_2A_2}=\frac{1}{2v}(m^2_{H_2}+2m^2_{A_2}), \notag \\
g_{H_2H_1^{\pm}H_1^{\mp}}=\frac{1}{v}(m^2_{H_2}+2m^2_{H_1^{\pm}}),& g_{H_2H_2^{\pm}H_2^{\mp}}=\frac{1}{v}(m^2_{H_2}+2m^2_{H_2^{\pm}}), &
g_{H_2H_2H_2H_1}=g_{H_1H_1H_1H_2}=0.\notag\\
\end{eqnarray}
From these expressions, a lower bound for all the scalar masses (other than $H_{1}$, which is always heavier than $H_2$ in this scenario),  can be set at $\gtrsim 63$ GeV, since there is no experimental observation of decays of the SM-Higgs boson to  other scalars.   This is in natural agreement with the current bounds for charged scalars, which set their masses above $\sim 80$ GeV \cite{Agashe:2014kda, Zyla:2020zbs}.  The recent signal for the rare three-body decay of the SM Higgs boson to photon and dileptons \cite{ATLAS:2021wwb}, will put extra constraints in the values of the allowed trilinear couplings.

The couplings of the gauge bosons to the SM Higgs have been determined with a $\sim 5\%$ precision \cite{Sirunyan:2018koj,Aad:2019mbh,ATLAS-CONF-2020-027}.  From our tree level expressions for  the gauge-Higgs couplings, Eqs.~(\ref{H12WW},\ref{H12ZZ}) we can parameterize  a deviation of the SM value by 
\begin{eqnarray}
\cos(\alpha - \theta) = \cos(\frac{\pi}{2} - \epsilon) = \sin\epsilon \equiv \delta ,
\end{eqnarray}
 where in the exact alignment limit $\delta=0 = \epsilon$.  A value of $\delta \lesssim 0.1$  is   compatible with the current experimental measurements and is consistent with our assumption of a 10\% deviation of the alignment limit in $(\alpha - \theta)$ in Fig.~\ref{figureAB-SM}. 

On the other hand, a deviation in the SM trilinear self-coupling $\lambda_{SM}$ will have an impact in di-Higgs production at tree-level \cite{Dawson:1998py,Frederix:2014hta}, single Higgs boson production and decays at one-loop level \cite{Degrassi:2016wml},  as well as in electroweak precision observables at two-loop level  \cite{Degrassi:2017ucl}.
In our case,  we can describe a small deviation of the alignment limit at tree level in terms of $\delta$, $\theta$ and $m_{h_0}$ as
\begin{eqnarray} 
g_{H_2H_2H_2} \equiv \lambda_{SM}\kappa_{\lambda} & = & \frac{m^2_{H_2}}{2v }\left[ (1+2\delta^2)\sqrt{1-\delta^2}  + \delta^3(\tan\theta -  \cot\theta)  - \frac{ m^2_{h_0}}{m^2_{H_2}} \frac{ \delta^3}{ 9 s_\theta c^3_\theta} \right]  , 
\label{eq:kappa}
\end{eqnarray}
where the term in square brackets $\kappa_{\lambda}$, is the scaling factor that parameterizes the deviation of  the SM Higgs trilinear self-coupling, in this case at tree level. The value of the trilinear self coupling has already been constrained experimentally \cite{CMS:2020tkr,ATLAS-CONF-2020-027}. In here,  we will make use of the modifier or $\kappa$ framework \cite{LHCHiggsCrossSectionWorkingGroup:2013rie} and the results in  \cite{Degrassi:2021uik}, where they set limits to  $\kappa_{\lambda}$, assuming the rest of the SM Higgs couplings to fermions and gauge bosons are the same or very close to the SM.  In our case, the value of $\kappa_{\lambda}$ will depend on $\delta$, $m_{h_0}$ and $\theta$. From Figure~\ref{figureAB-SM} we can see the dependence on $m_{h_0}$ on $\tan\theta$, which for a given $\delta$ allows us to determine the value of $\kappa_{\lambda}$.  As an example we take $\delta \sim 0.1$ and we fix $m_{h_0}$ to its possible maximum value for a given $\tan\theta$. In order to satisfy the bounds $-1.8 < \kappa_{\lambda} < 9.2$, as determined in \cite{Degrassi:2021uik}, $\tan\theta \lesssim 15$.  For smaller values of $\delta$ larger values of $\tan\theta$ are allowed. 

In case the alignment limit is exact, $\lambda_{SM}$ will still get corrections, but at loop level. In that case the factor $\kappa_{\lambda}$ will have a different expression, and depending on how complicated it is and what other restrictions are taken into account it might be possible to restrict the parameter space through it.

Analogous expressions for the couplings can be  found for scenario B. In this case, the SM-like Higgs boson would be $H_1$ and  the other neutral Higgs, $H_2$, would be lighter than the SM-like, at tree level.
As we already discussed, we cannot fully discard this possibility since in this alignment scenario, $H_2$ would not have couplings to the gauge bosons, and it could escape experimental detection.

We do not consider the most general case, without any alignment, since it implies that both neutral Higgs bosons couple to the gauge bosons, which is highly  restricted from the ex\-pe\-ri\-men\-tal data.

\subsection{Higgs one-loop self energy}
\label{loopHiggs-masses}

As we said before, the importance of having explicitly the Higgs couplings is relevant to calculate the radiative corrections or the possible loop contributions to different processes where the Higgs bosons are involved, including the radiative corrections to the SM Higgs mass and its renormalization \cite{Denner:1991kt}. For any BSM model the extra contributions to the oblique parameters \cite{Peskin:1991sw}, should fit the experimental data. There is work done in this direction for multi-Higgs models in  \cite{Grimus:2008nb,Hernandez:2015rfa}, where they explore the parameter space for N-Higgs doublet models. Their results imply that the Higgs masses should be almost degenerate, in a compact scalar spectrum. In their work, the assumption is that the new Higgs bosons mass scale should be above the EWSB scale, and that all scalars  couple to the gauge bosons. In our case, from the explicit form of the couplings, as e.g. in Eq.~(\ref{h0VV}), it can be seen that some Higgs loop contributions will not be present, so the relevant loop contribution calculated in \cite{Hernandez:2015rfa} will not appear for $h_0$, meaning that the restriction of $m_{h_0}>m_V$ (with $V=W, Z$) considered there is not required in our case. The same applies for the other Higgs bosons, (not the SM-like) considered in the two alignment scenarios that we explore  in this work, where some of the couplings to gauge bosons are null.

Considering CP-invariance, the renormalized neutral Higgs masses would be written as two dia\-go\-nal $3\times 3$ block matrices, one for the CP-even neutral states of the Higgs doublets $(h_{0}, H_{1}, H_{2})$, and the second for the CP-odd Higgs states $(A_{1}, A_{2}, G^{0})$
\begin{equation}
\mathcal{M}^{2}_{\phi^{0}}(s)=\mathcal{M}^{(0)2}_{H}+ 
\begin{pmatrix}
\hat{\Pi}^{S}(s) & 0 \\
0 & \hat{\Pi}^{P}(s) ~
\label{HmassBloq}
\end{pmatrix},
\end{equation}
where $\mathcal{M}^{(0)2}_{H}$ is the Higgs mass matrix at tree level given in section \ref{Hmasses}, the neutral part of expression (\ref{DiagHiggsMass}), with explicit tree-level neutral masses given in (\ref{h0masstree}), (\ref{H12masses}), (\ref{c3e44}) and (\ref{c3e45}).
The complete renormalized neutral Higgs self-energies
at one-loop level, $\hat{\Pi}^S(s)$ should be taken with the usual prescription, given for example  in \cite{Frank:2006yh}, adapting it to the S3-3H model. 

In our model, the one-loop contributions to the unrenormalized mass  corrections $\Pi^{S,(\tilde{q})}(s)$, that come only from the scalar sector self-energy,  denoted as $\Sigma^{\phi}$, would indicate corrections due to scalar bosons on the loop. In general we would have corrections to the Higgs masses considering scalar and gauge bosons, as well as fermions; in particular to the neutral scalar mass matrix we will have:

\begin{eqnarray}
\Pi^{S,(\tilde{q})}(s)=\Sigma^{\phi}(s)+\Sigma^{V}(s)+\Sigma^{f}(s) ~.
\end{eqnarray}

Due to the $\mathcal{Z}_2$ residual symmetry, the only trilinear coupling that involves a single $h_0$, which would give rise to a one-loop mass correction, is the one with two different charged scalars $h_0H_1^{\pm}H_2^{\mp}$ (Eq.~(\ref{h0Hc1Hc2}) in the Appendix). This mixed charged Higgs coupling is not present for the other two neutral Higgs bosons, avoiding the mixing of $h_0$ with the other neutral scalars at one-loop level, in this case via charged Higgs loops.

For the quartic couplings, there is no coupling that involves a single $h_0$ with a pair of identical Higgs bosons (including with three identical ones), see  Appendix \ref{AppA2}. This implies that there are no possible mass one-loop corrections that could mix $h_0$ with the other neutral scalars, $H_1$ and $H_2$. Moreover,  we can see from the gauge couplings with $h_0$ given in \ref{G-Hcoupl}, that there are only corrections to the $h_0$ mass but no mixing with other neutral Higgs bosons. Thus, the decoupling  is kept at one-loop level, as expected, with the consequence that the one-loop neutral scalar mass matrix will attain a block diagonal form 
\begin{eqnarray}
\Sigma^{\phi}(s)+\Sigma^{V}(s)=
\begin{pmatrix}
\Sigma^{\phi, V}_{h_0}(s) & 0& 0 \\
0 & \Sigma^{\phi, V}_{H_1}(s) & \Sigma^{\phi, V}_{H_1H_2}(s) \\
0 & \Sigma^{\phi, V}_{H_2H_1}(s) & \Sigma^{\phi, V}_{H_2}(s) \\
\end{pmatrix}.
\label{SmatxD} 
\end{eqnarray}
We see  from the above expression, that even at one-loop the $h_0$ scalar is decoupled from the other two, so the mass matrix structure of the other two neutral scalars is similar to the 2HDM.
 Nevertheless, there will be loop corrections to the $H_{1,2}$ masses due to $h_0$, as can be seen from the couplings (\ref{h0h0H1}) and (\ref{h0h0H2}).
On the other hand, $h_0$ will also receive corrections to its mass via the gauge boson loop, due to the allowed couplings (\ref{h0h0VV}).

The general scalar and gauge bosons contributions to  the square mass terms for $H_2$ and $H_1$ are given as:
\begin{eqnarray}
 \Sigma^{\phi, V}_{H_{n}}&=&\sum_{i}\frac{g_{H_nH_n\phi^0_i\phi^0_i}}{16\pi^2}A0(m_{\phi^0_i}^2)
+\sum_{i,j}\frac{g^2_{H_n\phi^0_i\phi^0_j}}{8\pi^2}B0(p^2,m_{\phi^0_i}^2,m_{\phi^0_j}^2)
+\sum_{k}\frac{g^2_{H_n\phi_k^{\pm}\phi_k^{\mp}}}{8\pi^2}B0(p^2,m^2_{\phi_k^{\pm}},m^2_{\phi_k^{\pm}})
\notag\\
&+&\sum_{i}\frac{g_{H_nH_nV_iV_i}}{16\pi^2}A0(m_{V_i}^2)
+\sum_{i}\frac{g^2_{H_nV_iV_i}}{8\pi^2}B0(p^2,m_{V_i}^2,m_{V_i}^2) ,
\end{eqnarray} \label{correctionsmass}
with $n=1,2$.\footnote{The terms where gauge bosons are involved show only the coupling contributions, the actual calculation will have to involve the gauge fixing.}  For the mixing term $H_{12}$ we get
\begin{eqnarray}
 \Sigma^{\phi,V}_{H_{1}H_2}&=&\sum_{i}\frac{g_{H_1H_2\phi^0_i\phi^0_i}}{16\pi^2}A0(m_{\phi^0_i}^2)
+\sum_{i,j}\frac{g_{H_1\phi^0_i\phi^0_j}g_{H_2\phi^0_i\phi^0_j}}{8\pi^2}B0(p^2,m_{\phi^0_i}^2,m_{\phi^0_j}^2)\notag \\
&+&
\sum_{k}\frac{g_{H_1\phi_k^{\pm}\phi_k^{\mp}}g_{H_2\phi_k^{\pm}\phi_k^{\mp}}}{8\pi^2}B0(p^2,m^2_{\phi_k^{\pm}},m^2_{\phi_k^{\pm}})
+\sum_{i}\frac{g_{H_1V_iV_i}g_{H_2V_iV_i}}{8\pi^2}B0(p^2,m_{V_i}^2,m_{V_i}^2)\notag \\
&+&
\sum_{k}\frac{g_{H_1\phi_k^{\pm}W^{\mp}}g_{H_2\phi_l^{\pm}W^{\mp}}}{8\pi^2}B0(p^2,m_{\phi^{\pm}_l}^2,m_W^2),
\end{eqnarray} \label{correctionsmassB}
where  $\phi^0_{i(j)}=h_0,H_1, H_2, A_1,A_2, G^0$, $\phi_k^{\pm}=H_{1,2}^{\pm}, G^{\pm}$ 
and $V_i=W^{\pm},Z^0$. In these expressions, $A0$ and $B0$ are the Passarino-Veltman functions of the masses involved \cite{Passarino:1978jh}.  
The radiative contributions to the mixing of $\Sigma^{\phi,V}_{H_1H_2}(s)$ reduce when we apply the alignment limit. For scenario A, the couplings reduce such that the one-loop corrections to the mixing term are given as follows
\begin{eqnarray}
 \Sigma^{\phi}_{H_{1}H_2}&=&\sum_{i}\frac{g_{H_1H_2\phi^0_i\phi^0_i}}{16\pi^2}A0(m_{\phi^0_i}^2)
+\sum_{i}\frac{g_{H_1\phi^0_i\phi^0_i}g_{H_2\phi^0_i\phi^0_i}}{8\pi^2}B0(p^2,m_{\phi^0_i}^2,m_{\phi^0_j}^2)\notag\\
&+&\sum_{k}\frac{g_{H_1\phi_k^{\pm}\phi_k^{\mp}}g_{H_2\phi_k^{\pm}\phi_k^{\mp}}}{8\pi^2}B0(p^2,m^2_{\phi_k^{\pm}},m^2_{\phi_k^{\pm}}),
\end{eqnarray} \label{correctionsmassA}
in this case we will only have  $\phi^0_{i}=h_0, A_1,A_2$, $\phi_k^{\pm}=H_{1,2}^{\pm}$, since all the terms involving gauge and Goldstone bosons vanish, so we simplify the notation to $\Sigma^{\phi}_{H_{1}H_2}$. This is taking  into account only the scalar and gauge contributions to the one-loop corrections. An equivalent expression can be found  for scenario B.

  We explore the structure of the loop contributions to the $h_0, H_1$ and $H_2$ masses coming from  the gauge bosons and  all scalars (neutral scalars and pseudoscalars, and charged scalars) by fixing the tree level masses and varying the $\theta$ parameter. 

A mixing term for $H_1$ and $H_2$ in the mass matrix, Eq.~(\ref{SmatxD}), would imply that they are not the true eigenstates. At one-loop level, we would expect this mixing parameter to be small, as the tree level should be the dominant order. Formally, we should take the poles of the propagator of the mass matrix at the order we are calculating to obtain the masses of the particles, in order to define two different states the mass matrix should be diagonalized at n-loop order [112]. Keeping this corrections small,  is another condition we could consider to constrain the free parameters of the model. Although complete NLO corrections should be taken into account (including fermions), our goal here is to show the importance of having the explicit form of the cubic and quartic couplings in order to be able to calculate loop corrections, which are functions of the model's parameters.

In Table \ref{mixingloop0} we present two example sets of parameters for these  corrections under scenario A. The choice of the benchmark points where these mixing parameters vanish, is meant to exemplify that there are regions of parameter space where indeed these one-loop corrections may be small. These examples correspond to points in parameter space where the mixing term in the mass matrix Eq. (\ref{SmatxD}) vanishes.

For the  light scalar spectrum  we achieve $\Sigma^{\phi}_{H_1H_2}(s)=0$ with $\tan\theta=1$,  but the scalar masses  are different, so the condition to keep the contributions to the oblique parameters $S,T$, small  might not be met. On the other hand, we find a spectrum with heavier scalar masses, where $\Sigma^{\phi}_{H_{1,2}}(s)=0$ and $\tan\theta\sim 2$.  A complete numerical exploration of these corrections could restrict more the parameter space, as they should be kept small. Our goal here is only to show the possible loop corrections that will be present in their general form. The small value of $\tan\theta$ found in these two examples indicates a maximal mixing between the $S_3$ singlet and doublet (see Eqs.(\ref{mh0pi4},\ref{mh12pi4})). It is also consistent with a numerical study of the S3-4H model (basically the S3-3H with one extra inert Higgs doublet), where compliance with the experimental Higgs bounds  was found for small values of $\tan\theta$, assuming  certain conditions on the Yukawa couplings \cite{Espinoza:2018itz}.

\begin{table}[tbp!]
    \centering
    \begin{tabular}{|c|c|c|}
    \hline
Scalar benchmarks  &  Masses (GeV) & $\tan\theta$ \\
    \hline
light spectrum &    $m_{h_0}=$ 80, $m_{H_1}=$ 200, $m_{A_{1,2}}=$ 80, $m_{H^{\pm}_{1,2}}=$ 100 & $ 1$ \\
    \hline
heavy spectrum &
    $m_{h_0}=$ 800, $m_{H_1}=$ 800, $m_{A_{1,2}}=$ 800, $m_{H^{\pm}_{1,2}}=$ 800 & $ 2.1$ \\
    \hline
\end{tabular}
\caption{Parameter values in scenario A that make the one-loop mixing parameter vanish, $\Sigma^{\phi}_{H_{1}H_2}=0$, taking into account only the scalar and gauge contributions.}
\label{mixingloop0}
\end{table}

Results in Table \ref{mixingloop0} are not conclusive, since we should take into account the fermionic contributions to have a more accurate estimation of the radiative corrections to scalar masses.
In particular, the top quark contribution is expected to be sizeable, due to its large Yukawa coupling. Work along these lines is in progress.

\section{Summary and Conclusions}
\label{Conclu}

The S3-3H model is an interesting and promising extension of the SM, that can accommodate well the masses and mixing of the quarks, leading to the NNI matrices \cite{Canales:2013cga},  as well as leptons, where it naturally gives a non-zero reactor mixing angle \cite{Mondragon:2007af,Canales:2012dr}. In this paper we studied the gauge and scalar sectors of this model. We chose a geometrical parameterization in spherical coordinates, which allowed us to express our results and analytical expressions in terms of the mixing angle, $\tan\theta$, between fields in the doublet and  the singlet irreps. From here it became clear that, in order to have realistic physical scenarios, without massless scalars,  this mixing must be always different from zero. As previously found \cite{Das:2014fea}, there is a residual $\mathcal{Z}_2$ symmetry, which decouples one of the neutral scalars $h_0$ from the gauge bosons.
This raises the interesting possibility of treating this decoupled scalar as a dark matter candidate, although we still have to probe its fermionic couplings. This possibility will be explored in a forthcoming publication. 

We performed a numerical analysis on the parameter space, taking into account unitarity and stability bounds, as well as the current experimental bounds on the charged masses. We studied two possible alignment scenarios, A and B, in which  one of the two  $\mathcal{Z}_2$ even neutral scalars $H_{1,2}$, is maximally coupled to the gauge bosons, and is thus taken to be as the SM Higgs. 
In scenario A, the lighter of the two neutral scalars, $H_2$, is the SM-like Higgs boson. The other possibility, scenario B, where the heavier $H_1$ is the SM-like Higgs boson, cannot be a priori excluded, since $H_2$ could have escaped detection due to the absence of couplings to the vector bosons.

We found the allowed ranges for the scalar masses, in terms of $\tan\theta$ in each alignment scenario, with a $10\%$ and $1\%$ precision on the $(\alpha-\theta)$ values. Our results show a clear restriction for all the scalar masses, which are mostly below TeV.  This corroborates similar analysis in this direction for scenario A, given in \cite{Das:2014fea} (although we allowed for a small deviation of the alignment limit, and for some uncertainty in the SM Higgs mass).  Scenario B in this model has not been analysed before.  The light $H_2$ scalar in this scenario opens the possibility that it might be regarded as the 96 GeV  scalar, which was suggested as a diphoton signal reported by CMS \cite{CMS-PAS-HIG-14-037}, and discussed in the literature in the context of SUSY and 2HDM models \cite{Heinemeyer:2018wzl,Haisch:2017gql,Biekotter:2019mib}. The same is possible for $h_0$ in both alignment limits.

We calculated all trilinear and quartic couplings between the Higgs bosons, and also among the Higgs and gauge bosons, giving analytical expressions in terms of the physical parameters of the model.  We found discrepancies with previously reported trilinear scalar couplings in this model  given in \cite{Barradas-Guevara:2014yoa},  where the $\mathcal{Z}_2$ symmetry is not reported nor explicitly present.  On the contrary,  our expressions do confirm the existence of the residual $\mathcal{Z}_2$ symmetry, as only $\Zd$ preserving couplings are present, consistent with the model structure reported in  \cite{Das:2014fea}, although they do not give the couplings explicitly. In our expressions, both the trilinear and quartic couplings for the SM-like Higgs boson reduce to the SM ones in the exact alignment limits. 

From our numerical analysis we found that the scalar masses  could be very close or even degenerate, but since we performed a random scan of the Higgs self-couplings, the allowed  masses shown do not imply they are necessarily degenerate for the same set of parameters, although we do find specific examples where this is the case. One such example is the heavy spectrum of Table  \ref{mixingloop0}, where all the scalar masses are degenerate, as is required to keep the contributions to the oblique parameters small.

The small deviation $\delta$ we considered of the alignment limit at tree level, is compatible with the latest experimental results on Higgs-gauge boson couplings. This deviation can also be used to  parameterize the contribution of the extra scalars of our model to the SM trilinear coupling $\lambda_{SM}$.  The current fits on the $\lambda_{SM}$ value, and our assumption of a deviation of the alignment limit of $\delta\sim 0.1$, sets an upper bound on $\tan\theta \lesssim 15$. 

In the exact alignment limit, some of the trilinear couplings depend only on the scalar masses, which sets a natural lower bound for all masses (other than $H_{1,2}$) to $\gtrsim 63$ GeV, since no SM Higgs decay into two lighter scalars has been observed  experimentally.  The inclusion of radiative corrections might change these bounds.

We obtained the analytical expressions for the one-loop corrections to the SM-like Higgs, due to scalar and gauge bosons in the loop, and found that the decoupling of $h_0$ remains at one-loop level, as expected from a symmetry of the Lagrangian. From the reduced  expressions for the couplings in scenario A, we calculated the value of $\tan\theta$ for which the one-loop mixing of $H_1$ and $H_2$ vanishes, $\Sigma^{\phi}_{H_{1}H_2}=0$, for two benchmark mass values. These results point to a value of $\tan\theta \approx \mathcal{O}(1)$, indicating a large mixing between the $S_3$ doublet and the singlet, consistent with what was reported in  \cite{Espinoza:2018itz}. 

The model  has different one-loop couplings to the  Higgs bosons through the extra $\Zd$ odd particles ($A_1,H^\pm_1,h_0$), as compared to the 2HDM.
Thus, although it reduces to a form similar to the 2HDM due to the presence of the residual $\mathcal{Z}_2$ symmetry, the extra $\Zd$ odd particles  will change the possible channels for decay and production of particles, and also the structure of radiative corrections. In particular they will have an impact on SM Higgs production, di-Higgs production, and loop corrections to the SM Higgs boson mass.

\section*{Acknowledgements}
We acknowledge useful discussions with Alexis Aguilar, Catalina Espinoza, and Genaro Toledo. This work was partially supported with  UNAM projects DGAPA PAPIIT IN111518 and IN109321. M.G.B would like to thank UDLAP for  financial support. A.P. acknowledges financial support from CONACyT, through grant 332430.

\appendix 
\section{Higgs Couplings } 
\label{appendixA}
\subsection{Scalar trilinear couplings}
\label{AppA1}
Here we explicitly write down the trilinear couplings of neutral with charged Higgs bosons and the rest of allowed quartic couplings of Higgs bosons. Here we are able to see the residual  $\mathcal{Z}_2$ symmetry,  the rest of the couplings are absent, as can be found from the direct calculation given in (\ref{HHH}) and (\ref{HHHH})

\begin{eqnarray}
g_{h_0H_1^\pm H_2^\pm} =  \frac{1}{3vs_\theta} \left( -m^2_{h_0} \frac{c_{2\theta} + c^2_\theta}{c_\theta} + 3m^2_{H_1^\pm}c_\theta - 3m^2_{H_2^\pm}c_\theta  \right),
\label{h0Hc1Hc2}
\end{eqnarray}

\begin{eqnarray}
g_{H_1H_1^\pm H_1^\pm} = \frac{2}{v s_{2\theta}} \left( -\frac{m^2_{h_0} }{3}s_{\alpha-\theta} + m^2_{H_1}c_{\theta}s_\alpha +2m^2_{H_1^\pm}c_{\theta}s_\alpha - 2m^2_{H_2^\pm} c^2_\theta s_{ \alpha- \theta}  \right),
\end{eqnarray}

\begin{eqnarray}
g_{H_2H^\pm_1H^\pm_1 }&=& \frac{2}{v s_{2\theta}}\left( 
\frac{m^2_{h_0} }{3}c_{\alpha -\theta} - m^2_{H_2}c_{\theta}c_\alpha  - 2m^2_{H^\pm_1}c_{\theta}c_\alpha  + 2m^2_{H^\pm_2} c^2_\theta c_{\alpha-\theta} \right) ,
\end{eqnarray}

\begin{eqnarray}
g_{H_1H_2^\pm H_2^\pm} = \frac{2}{v s_{2\theta}c^2_{\theta}} \left( \frac{m^2_{h_0}  }{9}s_{\alpha-\theta} + m^2_{H_1}c^2_\theta\left(s_\alpha c^3_\theta + c_\alpha s^3_\theta \right)  +m^2_{H_2^\pm}s_{2\theta}c^2_\theta c_{\alpha-\theta} \right) ,
\end{eqnarray}

\begin{eqnarray}
g_{H_2H^\pm_2 H^\pm_2} &=& \frac{2}{vs_{2\theta}} \left( -m^2_{h_0}\frac{c_{\alpha - \theta} }{9c^2_\theta} - m^2_{H_2} (c_{\alpha}c^3_\theta  - s_{\alpha} s^3_{\theta} )+ m^2_{H^\pm_2}s_{\alpha - \theta}s_{2\theta}  \right) , \label{c2e55}
\end{eqnarray}

\begin{eqnarray}
g_{H_2G^\pm G^\pm} = \frac{m^2_{H_2}s_{\alpha - \theta}}{v},
\end{eqnarray}
\begin{eqnarray}
g_{H_1G^\pm G^\pm} = \frac{m^2_{H_1}c_{\alpha - \theta}}{v},
\end{eqnarray}

\begin{eqnarray}
g_{h_0H_1^\pm G^\pm} = \frac{1}{v} (m^2_{h_0} - m^2_{H_1^\pm}),
\end{eqnarray}

\begin{eqnarray}
g_{H_2H_2^\pm G^\pm} = \frac{c_{\alpha-\theta}}{v} (m^2_{H_2} - m^2_{H_2^\pm}),
\end{eqnarray}

\begin{eqnarray}
g_{H_1H_2^\pm G^\pm} = \frac{s_{\alpha-\theta}}{v} (-m^2_{H_1} + m^2_{H_2^\pm}),
\end{eqnarray}

\begin{eqnarray}
g_{A_1 H^\pm_1 G^\mp} = \mp \frac{1}{v}( m^2_{H_1^\pm}- m^2_{A_1} +2c_\theta^2 ( m^2_{A_2} - m^2_{H_2^\pm})) ,
\end{eqnarray}

\begin{eqnarray}
g_{A_2 H^\pm_2 G^\mp} = \pm \frac{1}{v}(  m^2_{H_2^\pm}- m^2_{A_2} ) ,
\end{eqnarray}

\begin{eqnarray}
g_{A_1 H^\pm_1 H^\mp_2} = \pm \frac{1}{vt_\theta}( m^2_{H_1^\pm}- m^2_{A_1} +c_{2\theta} ( m^2_{A_2} - m^2_{H_2^\pm})) .
\end{eqnarray}

\subsection{Quartic scalar couplings}
\label{AppA2}

\begin{eqnarray}
\notag g_{H_2H_2H_1H_1} &=& \frac{1}{16v^2s^2_{2\theta}} \Bigg( \frac{4m^2_{h_0} s_{2(\alpha -  \theta)}}{3 c^2_\theta}\left( 2s_{2\alpha} +  s_{2(\alpha -  \theta)}  \right) - 2m^2_{H_1}s_{2\alpha}(3c_{2\alpha} s_{2(\alpha - \theta)}  \\
&&- 3s_{2\alpha} + s_{2\theta} ) + 2m^2_{H_2}s_{2\alpha}(3c_{2\alpha} s_{2(\alpha - \theta)} + 3s_{2\alpha} + s_{2\theta} ) \Bigg),  \label{c2e49}
\end{eqnarray}

\begin{eqnarray}
\notag g_{H_2H_2h_0h_0} &=& \frac{1}{2 v^2 s^2_{2\theta}} \left( m_{h_0}^2\left(c^2_{\alpha +\theta}-\frac{1}{3}c^2_{\alpha-\theta} \right)+ m^2_{H_1}c_{\theta}s_\alpha s_{2\alpha} c_{\alpha -\theta}  + m^2_{H_2}  c_\theta c_\alpha \left(c_{\alpha+\theta} +c_{2\alpha}c_{\alpha -\theta} \right)  \right),\\
\label{c2e51}
\end{eqnarray}

\begin{eqnarray}
g_{H_1H_1h_0h_0} = \frac{1}{2v^2s^2_{2\theta}} \left( m^2_{h_0}\left(s^2_{\alpha +\theta} -\frac{1}{3}s^2_{\alpha - \theta}\right)   + m^2_{H_1} c_\theta s_\alpha  \left(s_{\alpha +\theta} -c_{2\alpha} s_{\alpha-\theta} \right)  + m^2_{H_2}c_\theta c_\alpha  s_{2\alpha} s_{\alpha - \theta}   \right) ,\notag \\
\label{h0h0H1H1}
\end{eqnarray}

\begin{eqnarray}
g_{H_1H_2H_2H_2}= -\frac{1}{v^2s^2_{2\theta}} \left( \frac{2m^2_{h_0} c_\alpha c^2_{\alpha- \theta} (s_\alpha +2c_\theta s_{\alpha -\theta})}{9c^2_\theta  }  + \frac{m^2_{H_1} s^2_{2\alpha}s_{2(\alpha -\theta)} }{4} - \frac{m^2_{H_2} s_{2\alpha}c_{\alpha - \theta}(c_{\alpha - \theta}c_{2\alpha} + c_{\alpha +\theta}) }{2}  \right),\notag\\
\end{eqnarray}

\begin{eqnarray}
g_{H_1H_1H_1H_2}= -\frac{1}{v^2s^2_{2\theta}} \left( \frac{2m^2_{h_0} s_\alpha s^2_{\alpha -\theta}(c_\alpha +2c_\theta c_{\alpha -\theta}) }{9 c^2_\theta} +\frac{m^2_{H_2} s^2_{2\alpha}s_{2(\alpha -\theta)} }{4} +  \frac{m^2_{H_1}s_{2\alpha}s_{\alpha -\theta}( s_{\alpha -\theta}c_{2\alpha} - s_{\alpha +\theta})}{2}     \right),\notag\\
\end{eqnarray}

\begin{eqnarray}
g_{H_1H_2h_0h_0} = -\frac{1}{v^2s^2_{2\theta}} \left( \frac{m^2_{h_0}}{3}(2c_{2\alpha}s_{2\theta} +s_{2\alpha}c_{2\theta}) + m^2_{H_1}s_{2\alpha}s_{\alpha} c_{\theta} s_{\alpha -\theta} + m^2_{H_2}s_{2\alpha} c_\alpha c_{\theta} c_{\alpha - \theta}   \right), 
\end{eqnarray}

\begin{eqnarray}
\notag g_{H_2H_2A_1A_1} &=& \frac{1}{2v^2 s^2_\theta} \Bigg( -\frac{m^2_{h_0}}{36c^2_\theta}( 3c_{2\alpha} +3c_{2\theta}+4c^2_{\alpha -\theta}  ) + \frac{m^2_{H_1}s^2_\alpha c_\alpha c_{\alpha -\theta}}{2c_\theta} \nonumber \\
&&  +\frac{m^2_{H_2} c_\alpha}{2c_\theta}\left( c^2_\alpha c_{\alpha -\theta} - s_{\alpha}s_{\theta} \right)  + m^2_{A_1}c^2_\alpha - m^2_{A_2}(c^2_{\alpha} - s^2_{\theta}) \Bigg) , \label{c2e52}
\end{eqnarray}

\begin{eqnarray}
g_{H_1H_1A_1A_1} &=& \frac{1}{2v^2s^2_\theta} \Big( \frac{m^2_{h_0} }{36 c^2_\theta}(3c_{2\alpha}- 3c_{2\theta}  -4 s^2_{\alpha -\theta}  ) + \frac{m^2_{H_2} c^2_\alpha s_\alpha s_{\alpha - \theta}}{2c_\theta}   \nonumber \\ 
&&+  \frac{m^2_{H_1} s_\alpha  }{2 c_\theta }( s^2_\alpha s_{\alpha -\theta} +c_\alpha s_\theta)  +m^2_{A_1}s^2_\alpha + m^2_{A_2} (s^2_\theta - s^2_\alpha)   \Big) ,
\end{eqnarray}

{\small
\begin{eqnarray}
\notag g_{H_2H_2A_2A_2} &=& \frac{1}{2v^2s^2_{2\theta}} \Bigg( \frac{m^2_{h_0}}{9 c^2_\theta}(2c_{2\alpha}+s_{2\theta}s_{2(\alpha -\theta)} +  2c^2_{(\alpha -\theta)} +  2c_{2\theta} )
 \nonumber  \\
&& + \frac{m^2_{H_1}s_{2\alpha}}{4}\left(s_{2\alpha}+ s_{2\theta} + c_{2\theta}s_{2(\alpha -\theta)}  \right) + m^2_{A_2} s^2_{\alpha -\theta}s^2_{2\theta}\nonumber \\
&& + \frac{m^2_{H_2}}{8}(3c^2_{2\alpha} + 3c^2_{2\theta}+4c^4_{(\alpha -\theta)} + c_{2(\alpha -\theta)}  + 5 c_{2(\alpha +\theta)}  )  \Bigg) , \label{c2e53}
\end{eqnarray}}

\begin{eqnarray}
g_{H_1H_1A_2A_2} &=& \frac{1}{2v^2s^2_{2\theta}} \Bigg(- \frac{m^2_{h_0} }{9  c^2_\theta} (2c_{2\alpha} + s_{2\theta}s_{2(\alpha -\theta)} -2 s^2_{\alpha -\theta} -2c_{2\theta} ) \nonumber\\ &&+\frac{m^2_{H_2}s_{2\alpha}}{4}(s_{2\alpha} -s_{2\theta} +c_{2\theta}s_{2(\alpha -\theta)}) + m^2_{A_2}c^2_{\alpha - \theta}s^2_{2\theta}\nonumber \\   &&+  \frac{m^2_{H_1} }{8}(3c^2_{2\alpha} + 3c^2_{2\theta} - 4s^4_{\alpha - \theta} - c_{2(\alpha - \theta)} - 5c_{2(\alpha + \theta)} \Bigg),
\end{eqnarray}

\begin{eqnarray}
g_{h_0h_0A_1A_1}  =  \frac{m^2_{h_0} + 3 m^2_{H_1}s^2_\alpha + 3m^2_{H_2}c^2_\alpha }{12v^2 s^2_\theta},
\end{eqnarray}

\begin{eqnarray}
g_{h_0h_0A_2A_2}&=& \frac{1}{v^2s^2_{2\theta}} \Bigg( -\frac{m^2_{h_0} }{9}(4c^2_\theta +c_{2\theta}) + m^2_{H_1}c_\theta s_\alpha (s_{\alpha + \theta} -s_\theta c_\theta c_{\alpha -\theta})\nonumber \\
&& +m^2_{H_2}c_\theta c_\alpha (c_{\alpha + \theta} +s_\theta c_\theta s_{\alpha -\theta}) 
+ 2m^2_{A_1}c^4_\theta +2m^2_{A_2}c^2_\theta c_{2\theta}   \Bigg),\notag\\
\end{eqnarray}

\begin{eqnarray}
g_{H_1H_2A_1A_1} = \frac{1}{v^2 s^2_{2\theta}} \left(   \frac{m^2_{h_0}}{9}(3s_{2\alpha} + 2s_{2(\alpha -\theta)})  - m^2_{H_1} s_{2\alpha}s_\alpha c_\theta s_{\alpha - \theta}  - m^2_{H_2}s_{2\alpha} c_\alpha c_\theta c_{\alpha -\theta}    +  2 (m^2_{A_2}- m^2_{A_1})s_{2\alpha}c^2_{\theta}  \right) ,\notag\\
\end{eqnarray}

\begin{eqnarray}
g_{H_1H_2A_2A_2} &=& \frac{1}{2v^2s^2_{2\theta}} \Big( -\frac{2m^2_{h_0} }{9c^2_\theta}(s_{2\alpha} + 2c^2_{\theta}s_{2(\alpha - \theta)})   + \frac{m^2_{H_1}  }{2 } s_{2\alpha} (2c_{2\alpha} +s_{2\theta}s_{2(\alpha -\theta)} - 2 c_{2\theta})\nonumber \\  && - \frac{m^2_{H_2}  }{2}s_{2\alpha} (2c_{2\alpha} +s_{2\theta}s_{2(\alpha -\theta)} + 2 c_{2\theta})  + m^2_{A_2} s_{2(\alpha - \theta )}s^2_{2\theta}\Big),
\end{eqnarray}  

\begin{eqnarray}
g_{H_1h_0A_1A_2} = \frac{1}{v^2s^2_\theta} \left(  -\frac{m^2_{h_0}  }{9  c^2_\theta} (s_{\alpha +\theta} + c_\theta s_\alpha)(c_{2\theta} + c^2_\theta)  +  m^2_{H_1} s_\alpha c_\theta  - m^2_{H_2} (c_\alpha s_\theta + c^2_\theta s_{\alpha -\theta}) \right),
\end{eqnarray}

\begin{eqnarray}
g_{H_2h_0A_1A_2} = \frac{1}{v^2s^2_\theta} \left(  \frac{m^2_{h_0} }{9  c^2_\theta} (c_{\alpha +\theta} + c_\theta c_\alpha)(c_{2\theta} + c^2_\theta)   - m^2_{H_1} c_\alpha c_\theta  - m^2_{H_2} (s_\theta s_\alpha-c^2_\theta c_{\alpha -\theta} ) \right),
\end{eqnarray}

\begin{eqnarray}
g_{A_1A_1A_1A_1} =  \frac{m^2_{h_0} + 3 m^2_{H_1} s^2_\alpha + 3m^2_{H_2}c^2_\alpha }{24v^2 s^2_\theta },
\end{eqnarray}

\begin{eqnarray}
g_{A_2A_2A_2A_2} = \frac{1}{2v^2s^2_{2\theta}} \left( \frac{m^2_{h_0}}{9c^2_\theta}(2c^2_\theta + c_{2\theta}) + m^2_{H_1} (s_\alpha c_\theta - s^2_\theta s_{\alpha -\theta})^2 + m^2_{H_2}(c^2_\alpha c_{\alpha -\theta} -s_\alpha  s_\theta)^2 \right)  ,
\end{eqnarray}

\begin{eqnarray}
g_{A_1A_1A_2A_2}= \frac{1}{v^2s^2_{2\theta}} \left( \frac{m^2_{h_0} }{3} (c^2_{2\theta}-2c^2_\theta s^2_\theta) + m^2_{H_1}c_\theta s_\alpha(s_{\alpha +\theta} -s_\theta c_\theta c_{\alpha -\theta}) + m^2_{H_2} c_\theta c_\alpha(c_{\alpha +\theta} + s_\theta c_\theta s_{\alpha -\theta})  \right),\notag\\
\end{eqnarray}

\begin{eqnarray}
g_{H_1H_1G_0G_0} =  \frac{1}{2v^2} \left( \frac{m^2_{H_1}}{4s_{2\theta}} (s_{2\theta} + s_{2\alpha} - c_{2\alpha}s_{2(\alpha - \theta)}) - \frac{m^2_{H_2} s_{2\alpha} s^2_{\alpha -\theta}}{2s_{2\theta}}  + m^2_{A_2} s^2_{\alpha - \theta}  \right),
\end{eqnarray}

\begin{eqnarray}
g_{H_2H_2G_0G_0} =  \frac{1}{2v^2} \left( \frac{m^2_{H_1} s_{2\alpha} c^2_{\alpha -\theta}}{2s_{2\theta}} +  \frac{m^2_{H_2}}{4s_{2\theta}} (s_{2\theta} - s_{2\alpha} - c_{2\alpha}s_{2(\alpha - \theta)})   + m^2_{A_2} c^2_{\alpha - \theta}  \right),
\end{eqnarray}

\begin{eqnarray}
g_{H_1H_2G_0G_0} = -\frac{s_{2(\alpha -\theta)}}{4v^2s_{2\theta}} \left( (m^2_{H_1} - m^2_{H_2})s_{2\alpha} + 2m^2_{A_2} s_{2\theta}  \right),
\end{eqnarray}

\begin{eqnarray}
g_{G_0G_0A_2A_2} = \frac{1}{8v^2s_{2\theta}}\Big(m^2_{H_1}(3c_{2\theta}s_{2(\alpha -\theta)}-s_{2\alpha} + 3s_{2\theta})  -m^2_{H_2}(3c_{2\theta}s_{2(\alpha -\theta)}-s_{2\alpha} - 3s_{2\theta} ) \Big),
\end{eqnarray}

\begin{eqnarray}
g_{G_0G_0A_1A_1} = \frac{1}{2v^2} \left( m^2_{h_0} +\frac{m^2_{H_1} }{2s_\theta}s_\alpha c_{\alpha - \theta } - \frac{m^2_{H_2} }{2s_\theta} c_\alpha s_{\alpha -\theta}  \right),
\end{eqnarray}

\begin{eqnarray}
g_{H_1H_1A_2G_0} = - \frac{1}{2v^2 s_{2\theta}} \left(\frac{2m^2_{h_0} s^2_{\alpha -\theta}}{9 c^2_\theta} + \frac{m^2_{H_1}}{2}(2c_{2\theta } -s_{2\alpha}s_{2(\alpha -\theta)} -2c_{2\alpha}) + \frac{m^2_{H_2}}{2}s_{2\alpha} s_{2(\alpha - \theta )}   - m^2_{A_2}s_{2\theta} s_{2(\alpha - \theta)} \right),\notag\\
\end{eqnarray}

\begin{eqnarray}
g_{H_2H_2A_2G_0} = - \frac{1}{2v^2 s_{2\theta}} 
\left(\frac{2m^2_{h_0}c^2_{\alpha -\theta} }{9c^2_\theta}
+ \frac{m^2_{H_1}}{2} s_{2(\alpha - \theta )}
+ \frac{m^2_{H_2}}{2}(2c_{2\theta }-s_{2\alpha}s_{2(\alpha -\theta)} + 2c_{2\alpha})  + m^2_{A_2} s_{2\theta}s_{2(\alpha - \theta)} \right),\notag\\
\end{eqnarray}

\begin{eqnarray}
g_{H_1H_2A_2G_0} = \frac{1}{v^2 s_{2\theta} } \left( \frac{m^2_{h_0} s_{2(\alpha - \theta)}}{9 c^2_\theta} +  m^2_{H_1} s_{2\alpha} s^2_{\alpha -\theta} + m^2_{H_2} s_{2\alpha} c^2_{\alpha -\theta} - m^2_{A_2}c_{2(\alpha -\theta)}s_{2\theta}   \right),
\end{eqnarray}

\begin{eqnarray}
g_{H_1h_0A_1G_0}  =  \frac{1}{v^2s_\theta} \left( \frac{m^2_{h_0}}{3}\left( \frac{s_{\alpha + \theta}}{ c_\theta} + s_\alpha \right)  - m^2_{A_1}s_\alpha + m^2_{A_2}s_{\alpha -\theta}c_\theta \right),
\end{eqnarray}

\begin{eqnarray}
g_{H_2h_0A_1G_0} = \frac{1}{v^2s_\theta} \left( -\frac{m^2_{h_0}}{3}\left( \frac{c_{\alpha + \theta}}{ c_\theta} + c_\alpha \right)  + m^2_{A_1}c_\alpha - m^2_{A_2}c_{\alpha -\theta}c_\theta \right),
\end{eqnarray}

\begin{eqnarray}
g_{G_0A_2A_2A_2}= \frac{1}{v^2s_{2\theta}} \left(  -\frac{m^2_{h_0}}{9c^2_\theta} + \frac{m^2_{H_1} }{4}(2c_{2\alpha} +s_{2\theta} s_{2(\alpha - \theta)}-2c_{2\theta} ) -\frac{m^2_{H_2} }{4} (2c_{2\alpha} +s_{2\theta} s_{2(\alpha - \theta)}+2c_{2\theta} ) \right),\notag\\
\end{eqnarray}

\begin{eqnarray}
g_{G_0G_0G_0G_0} = \frac{m^2_{H_1}c^2_{\alpha -\theta} + m^2_{H_2} s^2_{\alpha - \theta}}{8v^2},
\end{eqnarray}

\begin{eqnarray}
g_{H_1G^0H^\pm_2 G^\mp} =  \pm \frac{1}{v^2}s_{\alpha - \theta} ( m^2_{H_2^\pm}- m^2_{A_2}),
\end{eqnarray}

\begin{eqnarray}
g_{H_2G^0H^\pm_2 G^\mp} =  \mp \frac{1}{v^2}c_{\alpha - \theta} (m^2_{H_2^\pm}- m^2_{A_2}),
\end{eqnarray}

\begin{eqnarray}
g_{H_1A_2H^\pm_2 G^\mp} =  \pm \frac{1}{v^2}c_{\alpha - \theta} (m^2_{H_2^\pm}- m^2_{A_2}),
\end{eqnarray}

\begin{eqnarray}
g_{H_2A_2H^\pm_2 G^\mp} =  \pm \frac{1}{v^2}s_{\alpha - \theta} (m^2_{H_2^\pm}- m^2_{A_2}),
\end{eqnarray}

\begin{eqnarray}
g_{h_0G^0H^\pm_1 G^\mp} = \mp \frac{1}{v^2}( m^2_{H_1^\pm}- m^2_{A_1} +2c_\theta^2 ( m^2_{A_2} - m^2_{H_2^\pm})),
\end{eqnarray}

\begin{eqnarray}
g_{h_0A_2H^\pm_1 G^\mp} =  \mp \frac{1}{v^2t_\theta}( m^2_{H_1^\pm}- m^2_{A_1} +c_{2\theta} ( m^2_{A_2} - m^2_{H_2^\pm})).
\end{eqnarray}

\subsection{Charged scalar-vector bosons couplings}
\label{AppA3}
For the couplings with charged Higgs bosons we have
\begin{eqnarray}
g_{H_1^\pm H_1^\pm W^\pm W^\mp} = \frac{2 M^2_W g^{\mu \nu}}{v^2} ,
\label{WWHcHc1}
\end{eqnarray}
\begin{eqnarray}
g_{H_2^\pm H_2^\pm W^\pm W^\mp} = \frac{2 M^2_W g^{\mu \nu}}{v^2} ,
\label{WWHcHc2}
\end{eqnarray}

\begin{eqnarray}
g_{H_1^\pm H_1^\pm ZZ}= \frac{g^2 \cos^22\theta_W  g^{\mu\nu}}{4\cos^2\theta_W},
\label{ZZHcHc1}
\end{eqnarray}
\begin{eqnarray}
g_{H_2^\pm H_2^\pm ZZ}= \frac{g^2 \cos^22\theta_W g^{\mu\nu}}{4\cos^2\theta_W},
\label{ZZHcHc2}
\end{eqnarray}

\begin{eqnarray}
g_{ H_1^\pm H_1^\mp \gamma \gamma} = e^2 g^{\mu\nu},
\end{eqnarray}
\begin{eqnarray}
g_{ H_2^\pm H_2^\mp \gamma \gamma} = e^2 g^{\mu\nu},
\end{eqnarray}
\begin{eqnarray}
g_{H_1^\pm H_1^\mp \gamma Z } = \frac{e g\cos 2\theta_W g^{\mu\nu} }{\cos\theta_W} ,
\end{eqnarray}
\begin{eqnarray}
g_{ H_2^\pm H_2^\mp \gamma Z} = \frac{e g\cos 2\theta_W g^{\mu\nu} }{\cos\theta_W}.
\end{eqnarray}

And the couplings for mixed charged and neutral Higgs bosons with gauge bosons, are given as

\begin{eqnarray}
g_{ H_2^\mp H_1 ZW^\pm} = \frac{g^{'2} \cos\theta_W \sin(\alpha -\theta)g^{\mu\nu}}{2}  ,
\end{eqnarray}
\begin{eqnarray}
g_{ H_2^\mp H_2 ZW^\pm} = -\frac{g^{'2} \cos\theta_W \cos(\alpha -\theta)g^{\mu\nu}}{2} ,
\end{eqnarray}
\begin{eqnarray}
g_{ H_2^\mp \gamma H_1 W^\pm} = -\frac{eg \sin(\alpha-\theta)g^{\mu\nu} }{2},
\end{eqnarray}
\begin{eqnarray}
g_{ H_2^\mp \gamma H_2 W^\pm} = \frac{eg \cos(\alpha-\theta)g^{\mu\nu}}{2},
\end{eqnarray}

\begin{eqnarray}
g_{\gamma G^0 W^\pm G^\mp} = \pm \frac{ge}{2}g^{\mu\nu},
\end{eqnarray}

\begin{eqnarray}
g_{Z G^0 W^\pm G^\mp} = \mp \frac{g^2s^2_{\theta_W}}{2c_{\theta_W}}g^{\mu\nu},
\end{eqnarray}

\begin{eqnarray}
g_{\gamma A_1 W^\pm H^\mp_1} = \pm \frac{ge}{2}g^{\mu\nu},
\end{eqnarray}

\begin{eqnarray}
g_{\gamma A_2 W^\pm H^\mp_2} = \pm \frac{ge}{2}g^{\mu\nu},
\end{eqnarray}

\begin{eqnarray}
g_{Z A_1 W^\pm H^\mp_1} = \mp \frac{g^2s^2_{\theta_W}}{2c_{\theta_W}}g^{\mu\nu},
\end{eqnarray}

\begin{eqnarray}
g_{Z A_2 W^\pm H^\mp_2} = \mp \frac{g^2s^2_{\theta_W}}{2c_{\theta_W}}g^{\mu\nu}.
\end{eqnarray}

The mixed charged Higgs boson and $h_0$ couplings with two gauge bosons are absent. 

\begin{eqnarray}
g_{\gamma H_1^+ H_1^+} = e(p+p')^\mu ,
\end{eqnarray}

\begin{eqnarray}
g_{\gamma H_2^+ H_2^+} = e(p+p')^\mu ,
\end{eqnarray}

\begin{eqnarray}
g_{H_1W^\pm H^\mp_2} =  \pm \frac{ig}{2} s_{(\alpha-\theta)}(p+p')^\mu,
\end{eqnarray}

\begin{eqnarray}
g_{H_2W^\pm H^\mp_2} =  \mp \frac{ig}{2} c_{(\alpha-\theta)}(p+p')^\mu,
\end{eqnarray}

\begin{eqnarray}
g_{h_0W^\pm H^\mp_1} =  \mp \frac{ig}{2}(p+p')^\mu, 
\end{eqnarray}

\begin{eqnarray}
g_{W^\pm H_2^\pm A_2} = \frac{g}{2}(p+p')^\mu  ,
\end{eqnarray}

\begin{eqnarray}
g_{W^\pm H_1^\pm A_1} = \frac{g}{2}(p + p' )^\mu ,
\end{eqnarray}

\begin{eqnarray}
g_{Z H_2 A_2}= \frac{g}{2\cos\theta_W}\cos(\alpha -\theta)(p+p')^\mu ,
\end{eqnarray}

\begin{eqnarray}
g_{Z H_1 A_2}= -\frac{g}{2\cos\theta_W}\sin(\alpha -\theta)(p+p')^\mu ,
\end{eqnarray}

\begin{eqnarray}
g_{H_1W^\pm G^\mp} =  \mp \frac{ig}{2} c_{(\alpha-\theta)}(p+p')^\mu,
\end{eqnarray}

\begin{eqnarray}
g_{H_2W^\pm G^\mp} =  \mp \frac{ig}{2} s_{(\alpha-\theta)}(p+p')^\mu,
\end{eqnarray}

\begin{eqnarray}
g_{W^\pm G^\pm G^0} = \frac{g}{2}(p+p')^\mu ,
\end{eqnarray}

\begin{eqnarray}
g_{\gamma G^+ G^+} = e(p+p' )^\mu .
\end{eqnarray}

\printbibliography
\end{document}